&pdflatex
\documentclass[a4paper,authoryear,3p]{elsarticle}

\usepackage{amssymb}
\usepackage{amsmath}
\usepackage{nicefrac}
\usepackage{tabu}

\begin{document}

\newcommand{\eqnspace}{1ex}
\newcommand{\matspace}{3pt}

\newcommand{\mspc}{\phantom{-}}

\let\originalleft\left
\let\originalright\right
\def\left#1{\mathopen{}\originalleft#1}
\def\right#1{\originalright#1\mathclose{}}

\begin{frontmatter}

\title{A WENO-type slope-limiter for a family of piecewise polynomial methods}

\author[mit,nasa]{Darren Engwirda\corref{cor1}}
\ead{engwirda@mit.edu}
\cortext[cor1]{Corresponding author\@. Tel.: +1-212-678-5521}

\author[nasa]{Maxwell Kelley}
\ead{maxwell.kelley@nasa.gov}


\address[mit]{Department of Earth, Atmospheric and Planetary Sciences, Massachusetts Institute of Technology, 54-918, 77 Massachusetts Avenue, Cambridge, MA 02139-4307, USA}

\address[nasa]{NASA Goddard Institute for Space Studies, 2880 Broadway, New York, NY 10025 USA}

\begin{abstract} 
A new, high-order slope-limiting procedure for the Piecewise Parabolic Method (PPM) and the Piecewise Quartic Method (PQM) is described. Following a Weighted Essentially Non-Oscillatory (WENO)-type paradigm, the proposed slope-limiter seeks to reconstruct smooth, non-oscillatory piecewise polynomial profiles as a non-linear combination of the \textit{natural} and \textit{monotone-limited} PPM and PQM interpolants. Compared to existing monotone slope-limiting techniques, this new strategy is designed to improve accuracy at smooth extrema, while controlling spurious oscillations in the neighbourhood of sharp features. Using the new slope-limited PPM and PQM interpolants, a high-order accurate Arbitrary-Lagrangian-Eulerian framework for advection-dominated flows is constructed, and its effectiveness is examined using a series of one- and two-dimensional benchmark cases. It is shown that the new WENO-type slope-limiting techniques offer a significant improvement in accuracy compared to existing strategies, allowing the PPM- and PQM-based schemes to achieve fully third- and fifth-order accurate convergence, respectively, for sufficiently smooth problems.


\end{abstract}


\end{frontmatter}

\section{Introduction}
\label{section_introduction}

Piecewise polynomial reconstruction is an important aspect of a vareity of high-order accurate numerical methods, providing a framework for (i) the evaluation of fluxes in finite-volume and discontinuous Galerkin schemes, and (ii) the construction of remapping operations for the Semi-Lagrangian and Arbitrary Lagrangian-Eulerian (ALE) methodologies. The performance of polynomial reconstruction is especially important in the context of advection-dominated transport, where the underlying dynamics support the development and propagation of sharp solution features, including shocks and rarefaction waves. Such systems can be modelled by the well-known conservative transport law
\begin{eqnarray}
\frac{\partial q}{\partial t} + \nabla\cdot\left(\mathbf{u}q\right) = 0,
\end{eqnarray}
where $q(\mathbf{x},t)$ is a conserved quantity and $\mathbf{u}(\mathbf{x},t)$ is a velocity field. In order to maintain stability in the presence of sharp and/or under-resolved solution features, polynomial reconstruction techniques are typically augmented by suitable `slope-limiting' procedures. Given a \textit{natural} polynomial reconstruction $Q(x)$, built as an interpolation of the discrete numerical solution, conventional slope-limiting procedures typically seek to limit the higher-order terms in the polynomial $Q(x)$ to ensure that the resulting profile is \textit{monotonicity-preserving}. Such reconstructions guarantee that the piecewise polynomial profile contains no new local extrema, and maintains oscillation-free behaviour as a result. While monotone slope-limiting strategies are known to result in robust and efficient numerical schemes, they tend also to artificially `flatten' local extrema, leading to a significant degradation in the overall accuracy of the scheme. This flattening effect manifests as spurious numerical dissipation, and can be a major source of error even when the underlying solution is sufficiently smooth and well-resolved.

High-order polynomial reconstruction plays a vital role in the formulation of conservative Semi-Lagrangian and Arbitrary Lagrangian-Eulerian (ALE) techniques. These methods play a critical role in the construction of contemporary global atmospheric and ocean circulation models -- being used to support generalised vertical coordinate models \citep{white2009high, bleck2002oceanic, halliwell2004evaluation} in which layer-wise atmospheric or oceanic dynamics are discretised using a direction-split ALE technique. Due to the long time integrations required by global climate modelling and numerical weather predication studies, a minimisation of numerical damping -- specifically, spurious diapycnal mixing -- is a key consideration. Following \citet{white2008high,white2009high}, a major motivation for the current study is to improve the accuracy of the underlying PPM and PQM reconstructions such that the level of artificial numerical dissipation induced by the action of the slope-limiter is minimised.

In additional to Semi-Lagrangian and Arbitrary Lagrangian-Eulerian schemes, polynomial reconstruction is also a key feature of Godunov-type \citep{godunov1959difference} finite-volume methods and discontinuous-Galerkin finite-element schemes \citep{reed1973triangularmesh,cockburn1998runge}, where it used to facilitate the evaluation of numerical fluxes. Considerable effort has been invested in the construction of higher-order accurate reconstruction techniques, starting with the second-order accurate Total Variation Diminishing (TVD) methods \citep{van1974towards,sweby1984high,barth1989limiter,leveque2002finite}, which are based on piecewise linear reconstructions, the third-order accurate Piecewise Parabolic Method (PPM) \citep{colella1984piecewise}, and, more recently, higher-order methods including the Parabolic and Quartic Spline Methods (PSM and QSM) \citep{zerroukat2006parabolic,Zerroukat20101150}, and the fifth-order accurate Piecewise Quartic Method (PQM) \citep{white2008high}. Such methods have all been designed to incorporate monotonicity-preserving slope-limiting strategies, and suffer a reduction in accuracy at smooth extrema as a result. 

The Weighted Essentially Non-Oscillatory WENO methods of \citet{liu1994weighted,shu1998essentially} are one of the few polynomial reconstruction schemes designed to offer genuinely higher-order accuracy, including the robust preservation of local extrema. Rather than seeking to preserve discrete monotonicity directly, such schemes are instead designed to achieve so-called \textit{non-oscillatory} reconstructions, in which the final polynomial profile for each grid-cell is calculated as a non-linear combination of a family of local candidate profiles. While the high-order accuracy and stability of such schemes is attractive, the requirement that multiple polynomial reconstructions be evaluated per grid-cell is a significant disadvantage -- leading to an appreciable increase in computational cost compared to the standard monotone piecewise polynomial methods. Additionally, the original WENO formulation was limited to the reconstruction of pointwise values along grid-cell boundaries and did not support the construction of full polynomial interpolants. Recent generalisations due to \citet{dumbser2007arbitrary} allow for WENO-like reconstructions of complete polynomials on arbitrary computational grids.

The present study is motivated by the desire to improve the performance of slope-limiting techniques for the PPM and PQM reconstruction schemes, ensuring that the accuracy of the polynomial interpolants are not degraded in the neighbourhood of well-resolved extrema. A conventional `two-pass' reconstruction procedure is investigated, in which an initial, unlimited polynomial reconstruction is first obtained, followed by a non-linear slope-limiting procedure. Rather than requiring the limited reconstruction to exactly satisfy discrete monotonicity constraints, locally non-monotone grid-cell profiles are instead replaced by a non-oscillatory reconstruction, generated using a variation of the non-linear WENO approach of \citet{liu1994weighted,shu1998essentially}. Compared to conventional techniques, it is shown that such a procedure does not lead to a reduction in the order-of-accuracy of the resulting numerical scheme when the underlying data is sufficiently smooth. Such behaviour is shown to significantly improve the accuracy of the high-order accurate PPM and PQM reconstruction methods.

The paper is organised as follows: in Sections~\ref{section_reconstruction},  \ref{section_edge_estimates} and \ref{section_monotone_limiter} the high-order Piecewise Parabolic and Piecewise Quartic methods (PPM and PQM) are reviewed, including a description of conventional monotonicity preserving slope-limiting strategies. The new WENO-type slope-limiting formulation is introduced in Section~\ref{section_weno_limiter}, detailing the use of optimal, non-linear weighting schemes. The resulting WENO-equipped PPM and PQM reconstruction techniques are used to formulate a high-order accurate one-dimensional semi-Lagrangian algorithm, with an extension to multi-dimensional problems achieved via a direction-splitting approach. These formulations are described in in Section~\ref{section_lagrangian}. Results for a series of one- and two-dimensional numerical experiments are presented in Section~\ref{section_results}, and the performance of the new polynomial interpolants and slope-limiting strategies is examined.

\section{Piecewise polynomial reconstruction}
\label{section_reconstruction}

\begin{figure}[t]

\label{figure_polyreco}

\begin{center}
\includegraphics[width=.425\textwidth]{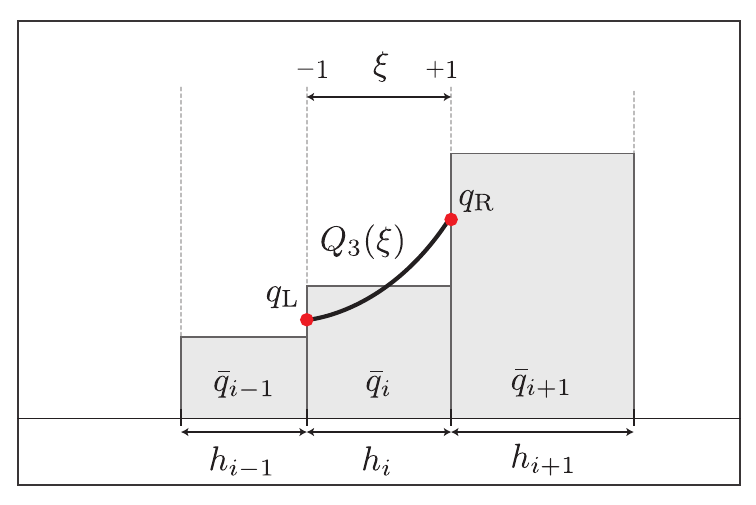}\quad
\includegraphics[width=.425\textwidth]{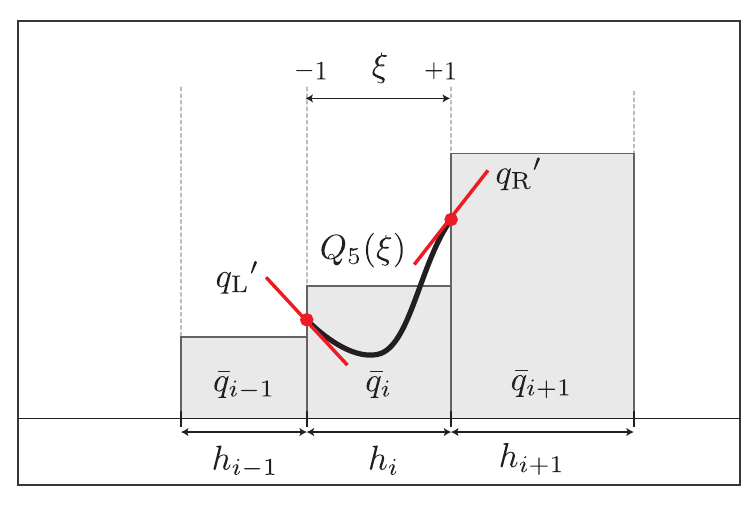}
\end{center}

\caption{Reconstruction of grid-cell polynomial $Q(\xi)$ from local data, showing (i) the piecewise parabolic method (PPM), and (ii) the piecewise quartic method (PQM).}

\end{figure}

Given a non-uniform, one-dimensional grid $\mathcal{G}$ spanning the points $\mathcal{X}=x_{1},x_{2},\dots,x_{n+1}$ and a set of discrete dynamical variables $\bar{Q} = \bar{q}_{1},\bar{q}_{2},\dots,\bar{q}_{n}$, the task is to \textit{reconstruct} a smooth piecewise polynomial interpolant $Q(x)$, over the grid-cells in $\mathcal{G}$. In this study, the reconstruction process is developed in the context of a finite-volume scheme, where the primary numerical variables are a set of discrete cell-mean quantities $\bar{q}_{i}$, defined such that
\begin{eqnarray}
\bar{q}_{i} = \frac{1}{h_{i}}\int_{x_\text{L}}^{x_{\text{R}}}\!Q_{i}(x)\,\mathrm{d}x,
\end{eqnarray} 
where $(x_{\text{L}},x_{\text{R}})$ are the left and right boundaries of a given grid-cell, and $h_{i} = x_{\text{R}}-x_{\text{L}}$ is the grid-cell width. Though $x$ is the primary underlying spatial coordinate, a local grid-cell coordinate $\xi$, defined in terms of the left and right grid-cell boundaries $x_{\text{L}}$ and $x_{\text{R}}$, such that
\begin{eqnarray}
x(\xi) = \tfrac{1}{2}(x_{\text{L}}+x_{\text{R}}) + \tfrac{1}{2}\xi(x_{\text{R}}-x_{\text{L}})
\end{eqnarray}
is used throughout, allowing the various interpolation schemes to be normalised onto grid-cells of varying widths. The local coordinate $\xi$ takes the values $\xi_{\text{L}}={-1}$ at the left grid-cell edge $x=x_{\text{L}}$ and $\xi_{\text{R}}={+1}$ at the right grid-cell edge $x=x_{\text{R}}$. In addition to theoretical convenience, this transformation is also known to improve the conditioning of the resulting numerical calculations. Based on considerations of global \textit{mass-conservation}, it is required that any reconstructed profile $Q(x)$ exactly satisfy local conservation constraints, such that 
\begin{eqnarray}
\label{equation_reconstruction_conservation}
\int_{h_{i}} Q_{i}\left(x\right)\,\mathrm{d}x = \int_{-1}^{+1} Q_{i}\left(\xi\right)\,\frac{\mathrm{d}x}{\mathrm{d}\xi}\,\mathrm{d}\xi = h_{i}\bar{q}_{i}.
\end{eqnarray}

\subsection{Piecewise parabolic reconstruction}

\medskip

The Piecewise Parabolic Method (PPM) \citep{colella1984piecewise} is a conservative, third-order accurate interpolant, based on the reconstruction of a set of local parabolic profiles
\begin{eqnarray}
\label{eqn_ppm}
Q(\xi) = \alpha_{0} + \alpha_{1}\xi + \alpha_{2}\xi^{2}.
\end{eqnarray}
The coefficients $\alpha_{i}\in\mathbb{R}$ can be found by requiring that (\ref{eqn_ppm}) conserve the local integral quantity $\bar{q}$, in addition to interpolating a pair of edge-value estimates $\left(q_{\text{L}},q_{\text{R}}\right)$, defined at the left/right grid-cell edges, respectively. These expressions can be written as a set of linear equations
\begin{eqnarray}
\tfrac{1}{2}{\displaystyle\int_{-1}^{+1}}Q(\xi)\,\mathrm{d}\xi = \bar{q},
\quad
Q(\xi_{\text{R}}) = q_{\text{R}},
\quad
Q(\xi_{\text{L}}) = q_{\text{L}}
\end{eqnarray}
the solution of which leads to explicit expressions for the coefficients $\alpha_{i}$
\begin{eqnarray}
\label{eqn_ppm_soln}
\begin{bmatrix}
\alpha_{0}\\[\matspace]
\alpha_{1}\\[\matspace]
\alpha_{2}
\end{bmatrix} = \mathbf{C}^{(3)}
\begin{bmatrix}
\bar{q}     \\[\matspace]
q_{\text{L}}\\[\matspace]
q_{\text{R}}
\end{bmatrix}\,,\quad\text{with}\quad
\mathbf{C}^{(3)} = 
\begin{bmatrix}
 \mspc\frac{3}{2}  & {-\frac{1}{4}}      & {-\frac{1}{4}} \\[\matspace]     
       \mspc 0     &   \mspc\frac{1}{2}  & {-\frac{1}{2}} \\[\matspace]
    {-\frac{3}{2}} &   \mspc\frac{3}{4}  &   \mspc\frac{3}{4}
\end{bmatrix}.
\end{eqnarray}
The PPM interpolant is completed through the selection of a suitable scheme to reconstruct local edge-value estimates $\left(q_{\text{L}},q_{\text{R}}\right)$. In this study, a family of explicit polynomial-based techniques is presented in Section~\ref{section_edge_estimates}. The PPM interpolant is illustrated in Figure~\ref{figure_polyreco}.

\subsection{Piecewise quartic reconstruction}

\medskip

The Piecewise Quartic Method (PQM) (\citet{white2008high}) is a conservative, fifth-order accurate interpolant, based on the reconstruction of a set of local quartic profiles
\begin{eqnarray}
\label{eqn_pqm}
Q(\xi) = \alpha_{0} + \alpha_{1}\xi + \alpha_{2}\xi^{2} + \alpha_{3}\xi^{3} + \alpha_{4}\xi^{4}.
\end{eqnarray}
The coefficients $\alpha_{i}\in\mathbb{R}$ can be found by requiring that (\ref{eqn_pqm}) conserve the local integral quantity $\bar{q}$, in addition to interpolating a pair of edge-value estimates $\left(q_{\text{L}},q_{\text{R}}\right)$, and edge-slope estimates $\left({q_{\text{L}}}',{q_{\text{R}}}'\right)$, defined at the left/right grid-cell edges, respectively. These expressions can be written as a set of linear equations
\begin{eqnarray}
\tfrac{1}{2}{\displaystyle\int_{-1}^{+1}}Q(\xi)\,\mathrm{d}\xi = \bar{q},
\quad
Q(\xi_{\text{R}}) = q_{\text{R}} ,
\quad
Q(\xi_{\text{L}}) = q_{\text{L}} ,
\quad
\frac{\partial Q}{\partial \xi}(\xi_{\text{R}}) = q_{\text{R}}',
\quad
\frac{\partial Q}{\partial \xi}(\xi_{\text{L}}) = q_{\text{L}}'
\end{eqnarray}
the solution of which leads to explicit expressions for the coefficients $\alpha_{i}$
\begin{eqnarray}
\label{eqn_pqm_soln}
\begin{bmatrix}
\alpha_{0} \\[\matspace]
\alpha_{1} \\[\matspace]
\alpha_{2} \\[\matspace]
\alpha_{3} \\[\matspace]
\alpha_{4}
\end{bmatrix} = \mathbf{C}^{(5)}
\begin{bmatrix}
\bar{q}\\
q_{\text{R}}  \\[\matspace]
q_{\text{L}}  \\[\matspace]
q_{\text{R}}' \\[\matspace]
q_{\text{L}}'
\end{bmatrix}\,,\quad\text{with}\quad
\mathbf{C}^{(5)} = 
\begin{bmatrix}
 \mspc\frac{15}{8}& -\frac{7}{16} & -\frac{7}{16} & \mspc\frac{1}{16}&-\frac{1}{16} \\[\matspace]
           \mspc0 &  \mspc\frac{3}{4}  & -\frac{3}{4}  &-\frac{1}{4} &-\frac{1}{4}  \\[\matspace]
-\frac{15}{4}& \mspc\frac{15}{8} &  \mspc\frac{15}{8} &-\frac{3}{8} & \mspc\frac{3}{8}  \\[\matspace]
           \mspc0 & -\frac{1}{4}  &  \mspc\frac{1}{4}  & \mspc\frac{1}{4} & \mspc\frac{1}{4}  \\[\matspace]
 \mspc\frac{15}{8}& -\frac{15}{16}& -\frac{15}{16}& \mspc\frac{5}{16}&-\frac{5}{16} 
\end{bmatrix}.
\end{eqnarray}
The PQM interpolant is completed through the selection of a suitable scheme to reconstruct local edge-value estimates $\left(q_{\text{L}},q_{\text{R}}\right)$ and edge-slope estimates $\left({q_{\text{L}}}',{q_{\text{R}}}'\right)$. In this study, a family of explicit polynomial-based techniques is presented in Section~\ref{section_edge_estimates}. The PQM interpolant is illustrated in Figure~\ref{figure_polyreco}.

\section{Edge estimates}
\label{section_edge_estimates}

The \textsc{ppm} and \textsc{pqm} interpolants described previously require that a set of edge-value and edge-slope estimates be computed at the left and right edges of each grid-cell in the mesh. Following \citet{colella1984piecewise} and \citet{white2008high}, such estimates can be computed via a secondary set of \textit{edge-centered} polynomial interpolants. In this study, such an approach is used to generate a family of high-order explicit schemes.

\begin{figure}[t]

\label{figure_polyedge}

\begin{center}
\includegraphics[width=.425\textwidth]{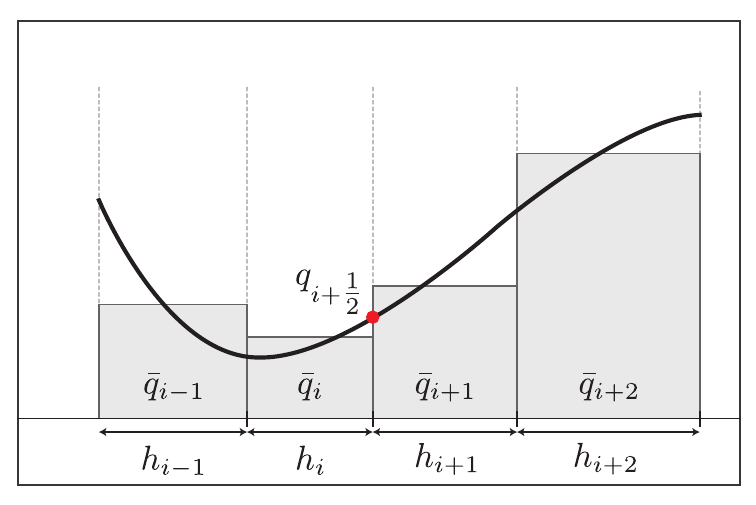}\quad
\includegraphics[width=.425\textwidth]{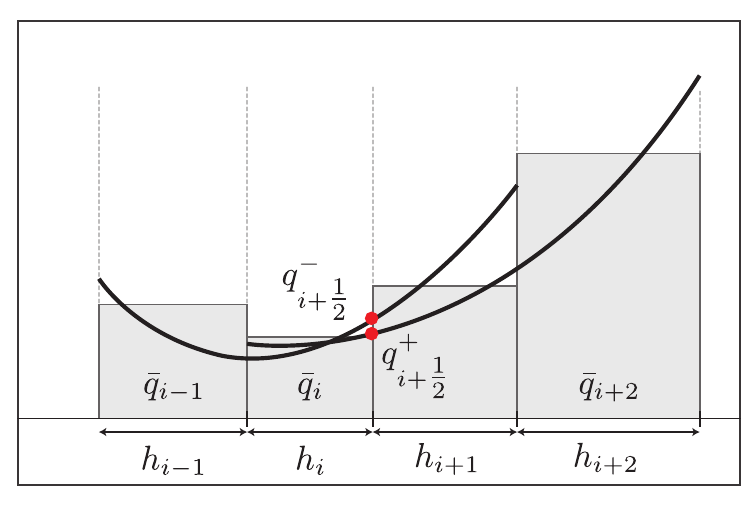}
\end{center}

\caption{Explicit edge reconstruction schemes, showing (i) a fourth-order accurate \textit{edge-centred} approximation for the edge-value at $q_{i+\frac{1}{2}}$ and (ii) the pair of left and right biased \textit{cell-centred} third-order accurate approximations to the edge-value $q_{i-\frac{1}{2}}^{-}$ and $q_{i-\frac{1}{2}}^{+}$.}

\end{figure}

\subsection{Explicit odd-degree schemes: P$_3$E and P$_5$E}

\medskip

A conservative, \textit{edge-centred}, polynomial interpolant
\begin{eqnarray}
\label{eqn_odd_edge}
Q(\Delta) = \alpha_{0} + \alpha_{1}\Delta + \dots + \alpha_{n}\Delta^{n-1}, \quad\text{with}\quad \Delta = x - x_{i+\frac{1}{2}}
\end{eqnarray}
can be defined about each interior edge $x_{i+\frac{1}{2}}$ in the mesh, by requiring that (\ref{eqn_odd_edge}) satisfy local conservation constraints over a stencil $\mathcal{S}_{i+\frac{1}{2}} = \left\{i-\frac{n}{2}+1,\dots,i,i+1,\dots,i+\frac{n}{2} \right\}$, where the set $\mathcal{S}_{i+\frac{1}{2}}$ is symmetric about the edge $i+\frac{1}{2}$. These constraints can be expressed as a set of linear equations
\begin{eqnarray}
\label{eqn_edge_odd_matrix}
{\displaystyle\int_{\Delta_{\text{L}_{j}}}^{\Delta_{\text{R}_{j}}}} Q(\Delta)\,\mathrm{d}\Delta 
=
{\displaystyle\int_{\Delta_{\text{L}_{j}}}^{\Delta_{\text{R}_{j}}}} \left[1, \Delta, \dots, \Delta^{n-1}\right]\,\mathrm{d}\Delta
\begin{bmatrix}
\alpha_{0} \\
\alpha_{1} \\
\vdots \\
\alpha_{n}
\end{bmatrix} =  
h_{j}\bar{q}_{j},
\qquad \forall j\in\mathcal{S}_{i+\frac{1}{2}}.
\end{eqnarray}
Once (\ref{eqn_edge_odd_matrix}) is assembled, it can be solved for the polynomial coefficients $\alpha_{i}$ in (\ref{eqn_odd_edge}). Noting that $\Delta=0$ at $x=x_{i+\frac{1}{2}}$, estimates for the edge-value and edge-slope at the edge $x_{i}$ can be computed through an evaluation of (\ref{eqn_odd_edge}) and its derivative, leading to
\begin{eqnarray}
q_{i+\frac{1}{2}} = \alpha_{0}, \quad\text{and}\quad \left(\frac{\partial q}{\partial x}\right)_{i+\frac{1}{2}} = \alpha_{1}.
\end{eqnarray}
Recalling that the PPM interpolant provides a locally third-order accurate reconstruction, a set of edge-value estimates of at least third-order accuracy are required to preserve the formal order-of-accuracy of the scheme. Following similar arguments, the PQM interpolant requires a set of fifth-order accurate edge-value and edge-slope estimates in order to preserve formal accuracy. In this study, a pair of methods, based on local cubic and quintic polynomials, are used to provide suitably high-order edge estimates for the PPM and PQM schemes. The P$_3$E scheme, based on a local cubic polynomial spanning the four grid-cells adjacent to a given edge, provides a fourth-order accurate approximation to the edge-values, and a third-order accurate approximation to the edge-slopes. The P$_5$E scheme, based on a local quintic polynomial spanning the six grid-cells adjacent to a given edge, provides a sixth-order accurate approximation to the edge-values and a fifth-order accurate approximation to the edge-slopes. The P$_3$E scheme exceeds the accuracy requirements of the PPM interpolant, while the P$_5$E scheme satisfies those of the PQM reconstruction. Given a uniform grid-spacing, solutions to (\ref{eqn_edge_odd_matrix}) lead to simple expressions for the coefficients $\alpha_{i}$. Explicit coefficients for the P$_2$E and P$_4$E schemes are provided in \ref{edge_appendix}. For non-uniform grid-spacing, (\ref{eqn_edge_odd_matrix}) can be factored via numerical techniques.

\subsection{Explicit even-degree schemes: P$_2$E and P$_4$E}

\medskip

A conservative, \textit{cell-centred}, polynomial interpolant
\begin{eqnarray}
\label{eqn_even_edge}
Q(\Delta) = \alpha_{0} + \alpha_{1}\Delta + \dots + \alpha_{n}\Delta^{n-1}, \quad\text{with}\quad \Delta = x - \tfrac{1}{2}\left(x_{\text{L}}+x_{\text{R}}\right)
\end{eqnarray}
can be defined about each interior grid-cell $i$ in the mesh, by requiring that (\ref{eqn_even_edge}) satisfy local conservation constraints over a stencil $\mathcal{S}_{i} = \left\{i-\lfloor\frac{n}{2}\rfloor,\dots,i-1,i,i+1,\dots,i+\lfloor\frac{n}{2}\rfloor \right\}$, where the set $\mathcal{S}_{i}$ is symmetric about the cell $i$. These constraints can be expressed as a set of linear equations
\begin{eqnarray}
\label{eqn_edge_even_matrix}
{\displaystyle\int_{\Delta_{\text{L}_{j}}}^{\Delta_{\text{R}_{j}}}} Q(\Delta)\,\mathrm{d}\Delta = 
{\displaystyle\int_{\Delta_{\text{L}_{j}}}^{\Delta_{\text{R}_{j}}}} \left[1, \Delta, \dots, \Delta^{n-1}\right]\,\mathrm{d}\Delta
\begin{bmatrix}
\alpha_{0} \\
\alpha_{1} \\
\vdots \\
\alpha_{n}
\end{bmatrix} = 
h_{j}\bar{q}_{j},
\qquad \forall j\in\mathcal{S}_{i}.
\end{eqnarray}
Once (\ref{eqn_edge_even_matrix}) is assembled, it can be solved for the polynomial coefficients $\alpha_{i}$ in (\ref{eqn_even_edge}). Given the cell-wise interpolants (\ref{eqn_even_edge}), estimates for the edge-values and edge slopes can be computed by evaluating (\ref{eqn_even_edge}) and its derivative at the grid-cell boundaries. Noting that two cell-wise interpolants are available for each interior edge, estimates are computed using a mean value
\begin{eqnarray}
q_{i+\frac{1}{2}} = \frac{1}{2}q^{-}_{i+\frac{1}{2}}+\frac{1}{2}q^{+}_{i+\frac{1}{2}},
\quad\text{and}\quad
\left(\frac{\partial q}{\partial x}\right)_{i+\frac{1}{2}}  = \frac{1}{2}\left(\frac{\partial q}{\partial x}\right)^{-}_{i+\frac{1}{2}}+\frac{1}{2}\left(\frac{\partial q}{\partial x}\right)^{+}_{i+\frac{1}{2}}
\end{eqnarray}
where the local coordinate $\Delta = x_{i+\frac{1}{2}} - \tfrac{1}{2}\left(x_{\text{L}}+x_{\text{R}}\right)^{\mp}$ is evaluated at the edge $x_{i+\frac{1}{2}}$ and $(\cdot)^{-},\,(\cdot)^{+}$ denote an evaluation of the neighbouring cell-centred interpolants located to the left and right of the given edge $x_{i+\frac{1}{2}}$ respectively. In this study, a pair of even-degree methods, based on local quadratic and quartic interpolants, are used to provide suitably high-order edge estimates for the PPM and PQM schemes. The P$_2$E scheme, based on a local quadratic polynomial spanning the three grid-cells adjacent to a given cell, provides a third-order accurate approximation to the edge-values, and a second-order accurate approximation to the edge-slopes. The P$_4$E scheme, based on a local quartic polynomial spanning the five grid-cells adjacent to a given cell, provides a fifth-order accurate approximation to the edge-values and a fourth-order accurate approximation to the edge-slopes. The P$_2$E scheme matches the accuracy requirements of the PPM interpolant exactly, while the P$_4$E scheme offers matching accuracy edge-value estimates for the PQM interpolant, and edge-slope estimates of one order lower than required. Additional discussion of the P$_2$E and P$_4$E schemes is presented in \ref{edge_appendix}.

\subsection{Edge estimates at domain boundaries}

\medskip

The P$_2$E, P$_3$E, P$_4$E and P$_5$E interpolants presented previously all rely on a symmetric stencil of neighbouring grid-cell values. In the vicinity of domain boundaries, such stencils do not exist. Following \citet{white2008high}, edge-estimates in such cases are computed using high-order \textit{one-sided} techniques, in which the value and slope of the nearest interior polynomial is extrapolated to the boundary edges. 

\section{Monotone slope-limiting}
\label{section_monotone_limiter}

The PPM and PQM interpolants presented in Section~\ref{section_reconstruction} are not automatically guaranteed to respect local monotonicity constraints, and may instead contain spurious oscillations and overshoots in the neighbourhood of any sharp and/or poorly resolved features present in the underlying data. It is therefore necessary to apply a suitable \textit{slope-limiting} procedure to the reconstructed profiles $Q(\xi)$, designed to modify the polynomial coefficients to ensure that the limited profiles are adequately bounded with respect to adjacent cell-mean values. In the following section, a pair of \textit{monotone} slope-limiters for the PPM and PQM interpolants are reviewed, designed to enforce exact cell-wise monotonicity within each grid-cell.

\subsection{Limiting PPM}

\medskip

Enforcing exact monotonicity for the PPM interpolant is a two-stage process, in which the edge-value estimates $q_{\text{L}}$ and $q_{\text{R}}$ for each grid-cell are modified to ensure that the cell-wise profiles $Q_{i}(\xi)$ are bounded by the set of neighbouring cell-mean values $\{\bar{q}_{i-1},\bar{q}_{i},\bar{q}_{i+1}\}$. In addition to suppressing spurious oscillations, such a limiter ensures that the resulting PPM profiles obey a discrete maximum principle, and do not introduce any new extrema in the underlying data. Following \citet{colella1984piecewise}, the action of the slope-limiter is accomplished in two stages. Firstly the boundedness of the edge-value estimates is checked and enforced, ensuring that each edge-value estimate is consistent with the neighbouring cell-mean quantities. Secondly, the consistency of the cell-wise PPM profiles are themselves checked, and are modified to ensure that they respect adjacent cell-mean values. This process can result in further modifications to the edge-value estimates. The monotone limiter for the PPM reconstruction is described in detail in \citet{colella1984piecewise} and is summarised in \ref{ppm_poly_appendix}.

\subsection{Limiting PQM}

\medskip

Consistent with the strategy discussed previously for PPM, the enforcement of monotonicity constraints for the PQM reconstruction is again realised as a two-stage process, in which the edge-value and edge-slope estimates $\left(q_{\text{L}},{q_{\text{L}}}'\right)$ and $\left(q_{\text{R}},{q_{\text{R}}}'\right)$ are first modified to ensure that they respect the local distribution of cell-mean data, followed by modifications to the grid-cell profiles $Q_{i}(\xi)$ themselves. The resulting monotone PQM reconstruction is guaranteed to suppress spurious oscillations and to be free of new local extrema. Following \citet{white2008high}, the action of the slope-limiter is accomplished in multiple stages. Firstly the boundedness of the edge-value estimates is checked and enforced, ensuring that each edge-value estimate is consistent with the neighbouring cell-mean quantities. Secondly, the consistency of the edge-slope estimates are evaluated, with the slopes modified to ensure that they are in agreement with a local linear estimate. Finally, the consistency of the cell-wise PQM profiles are themselves checked, and are modified to ensure that they respect the adjacent cell-mean values. This process can result in further modifications to both the edge-slope and edge-value estimates. The monotone limiter for the PPM reconstruction is described in detail in \citet{white2008high} and is summarised in \ref{pqm_poly_appendix}.

\section{Non-oscillatory slope-limiting strategies}
\label{section_weno_limiter}

While the conventional \textit{monotonicity-preserving} techniques described in Section~\ref{section_monotone_limiter} result in robust and oscillation-free reconstructions, it is well-known \citep{white2008high,Zerroukat20101150} that such slope-limiters can seriously compromise the accuracy of the underlying high-order schemes. Recalling that all local extrema are explicitly \textit{flattened} by such methods, it is clear that monotonicity-preserving schemes reduce to low-order representations in the neighbourhood of such features -- even when they are sufficiently smooth and well-resolved. While this degradation in accuracy is initially restricted to the grid-cells immediately adjacent to such features, the associated numerical dissipation over time can lead to non-local diffusive errors. It is clear that such behaviour is highly undesirable in the construction of low-dissipation, high-order accurate numerical methods for transport phenomena. \citet{white2008high} report that the use of monotonicity-preserving slope-limiters in a PPM- and PQM-based Aribtrary Lagrangian-Eulerian remapping algorithm reduced the global order-of-accuracy of the schemes to second-order for a range of one-dimensional test problems.

In this section, an alternative slope-limiting technique for both the PPM and PQM reconstructions are described. Drawing on the well-known Weighted Essentially Non-Oscillatory (WENO) methodology, originally introduced in \citet{liu1994weighted,shu1998essentially}, the alternative slope-limiter seeks to build a locally smooth, non-oscillatory polynomial interpolant as a non-linear combination of local profiles. Importantly, such an approach does not require that the piecewise interpolants be exactly monotone, but instead aims to preserve the high-order accuracy of the underlying scheme when the data is smooth and well-resolved, while also controlling spurious oscillations near sharp and/or under-resolved features. Compared to conventional WENO schemes, the approach presented here does not require the computation of multiple polynomial reconstructions for each grid-cell, but instead relies on the local profiles that arise naturally in the PPM and PQM schemes. Such a process greatly improves computational efficiency.

\subsection{A WENO-type slope-limiter}

\medskip

Based on a WENO-type philosophy \citep{dumbser2007arbitrary}, it is proposed that a smooth, essentially non-oscillatory polynomial reconstruction be obtained within each grid-cell as a non-linear convex combination of the \textit{natural} and \textit{monotone-limited} PPM or PQM polynomials
\begin{eqnarray}
\label{eqn_weno_blend}
Q_{i}(x) = \hat{w}_{i,n}\hat{Q}_{i,n}(x) + \hat{w}_{i,m}\hat{Q}_{i,m}(x).
\end{eqnarray}
Here $Q_{i}(x)$ is the final, non-oscillatory polynomial reconstruction for a given grid-cell, $\hat{Q}_{i,n}(x)$ and $\hat{Q}_{i,m}(x)$ are the so-called \textit{natural} and \textit{monotone-limited} polynomial reconstructions associated with the same grid-cell, and $\hat{w}_{i,n},\hat{w}_{i,m}\in\mathbb{R}^{+}$ are a pair of \textit{non-linear} weights, defined such that $\hat{w}_{i,n}+\hat{w}_{i,m}=1$. The natural interpolant $\hat{Q}_{i,n}(x)$ is simply the unlimited polynomial profile that is obtained from the unmodified PPM or PQM reconstruction. The monotone-limited profiles $\hat{Q}_{i,m}(x)$ are those produced by the monotone slope-limiting strategies presented in Section~\ref{section_monotone_limiter}. The non-linear combination (\ref{eqn_weno_blend}) defines a \textit{blending} between the natural and monotone-limited grid-cell polynomial profiles. The idea of blending whole grid-cell profiles according to WENO-like weights is adapted from \citet{dumbser2007arbitrary}, in which a WENO-type scheme was developed for unstructured computational grids. The use of related methods for the Piecewise Parabolic Method has previously been investigated by \citet{blossey2008selective}. 

An \textit{optimal} slope-limiting strategy can be developed using (\ref{eqn_weno_blend}) by noting that: (i) the natural interpolant $\hat{Q}_{i,n}(x)$ automatically provides a full-order accurate reconstruction, including at smooth extrema, and (ii) the monotone-limited profile $\hat{Q}_{i,m}(x)$ provides a non-oscillatory representation about sharp and/or under-resolved regions. Given such behaviour, the task is to define the pair of non-linear weights $\hat{w}_{i,n}$ and $\hat{w}_{i,m}$ such that $\hat{w}_{i,n}\rightarrow 1$ and $\hat{w}_{i,m}\rightarrow 0$  when the underlying data is sufficiently smooth and, conversely, that $\hat{w}_{i,n}\rightarrow 0$ and $\hat{w}_{i,m}\rightarrow 1$ for grid-cells that lie in the neighbourhood of discontinuous features. Such a weighting scheme can be realised through the use of an \textit{oscillation-indicator} -- a scalar value computed for each grid-cell in the mesh that is designed to provide a measure of the relative smoothness of the local distribution of cell-mean data $\bar{q}_{i}$. In the standard WENOframework \citep{shu1998essentially}, oscillation indicators are defined as a function of the higher-order derivative terms associated with a particular grid-cell reconstruction $Q_{i,j}(x)$, such that
\begin{eqnarray}
\label{eqn_oscillation_indicator}
\beta_{i,j} = \sum_{m=1}^{d} \int_{h_{i}} h_{i}^{2m-1} \left(\frac{\partial^{m} Q_{i,j}}{\partial x^{m}}\right)^{2}\,\mathrm{d}x,
\end{eqnarray}
where the factors $h_{i}^{\alpha}$ are included to ensure that (\ref{eqn_oscillation_indicator}) is scale independent. Grid-cells that contain discontinuous and/or poorly-resolved data have $\beta_{i,j}\gg 1$. Following \cite{blossey2008selective}, an approximation to the integral oscillation indicators is used in this study to improve computational efficiency, such that
\begin{equation}
\label{eqn_oscl}
\beta_{i,j} = 
\left(h_{i}\frac{\partial \pi_{j}}{\partial x}(m_{i})\right)^{2} +
\left(h_{i}^{2}\frac{\partial^{2} \pi_{j}}{\partial x^{2}}(m_{i})\right)^{2},
\end{equation}
where $h_{i}$ is the width of the $i$-th grid cell, $x=m_{i}$ is the grid-cell midpoint and $\pi_{j}(x)$ is a local quadratic approximation $\pi_{j}(x)=\hat{\pi}_{0}+\hat{\pi}_{1}x+\hat{\pi}_{2}x^2$, defined for each grid-cell in the mesh. The quadratic profile is computed as a local conservative interpolant
\begin{eqnarray}
{\displaystyle\int_{h_{j}}}\pi_{i}(x)\,\mathrm{d}x= 
{\displaystyle\int_{h_{j}}}[1,x,x^{2}]\,\mathrm{d}x\, 
\begin{bmatrix}
\hat{\pi}_{0} \\
\hat{\pi}_{1} \\
\hat{\pi}_{2}
\end{bmatrix} = 
h_{j}\bar{q}_{j},\qquad \text{for } j = i-1,\dots,i+1.
\end{eqnarray}
Note that in (\ref{eqn_oscl}), computation of the oscillation indicator $\beta_{i,j}$ requires the evaluation of the derivatives of each of the $j\in\mathcal{S}_{i}$ polynomials $\pi_{j}(x)$ associated with the stencil of the $i$-th grid-cell. The first- and second-derivatives of the polynomials $\pi_{j}(x)$ only need to be computed \textit{once} for each grid-cell and then re-used in the calculation of the various neighbouring $\beta_{i,j}$. In this way, the functions $\pi_{j}(x)$ do not need to be re-computed multiple times. The use of the secondary reconstruction $\pi_{i}(x)$ to locate discontinuities in the underlying data is related to so-called \textit{troubled-cell} techniques, previously explored by a range of authors including \citet{balsara2007sub} and \citet{qiu2005comparison}.

\begin{figure}[t]

\label{figure_weno_weights}

\begin{center}
\includegraphics[width=.60\textwidth]{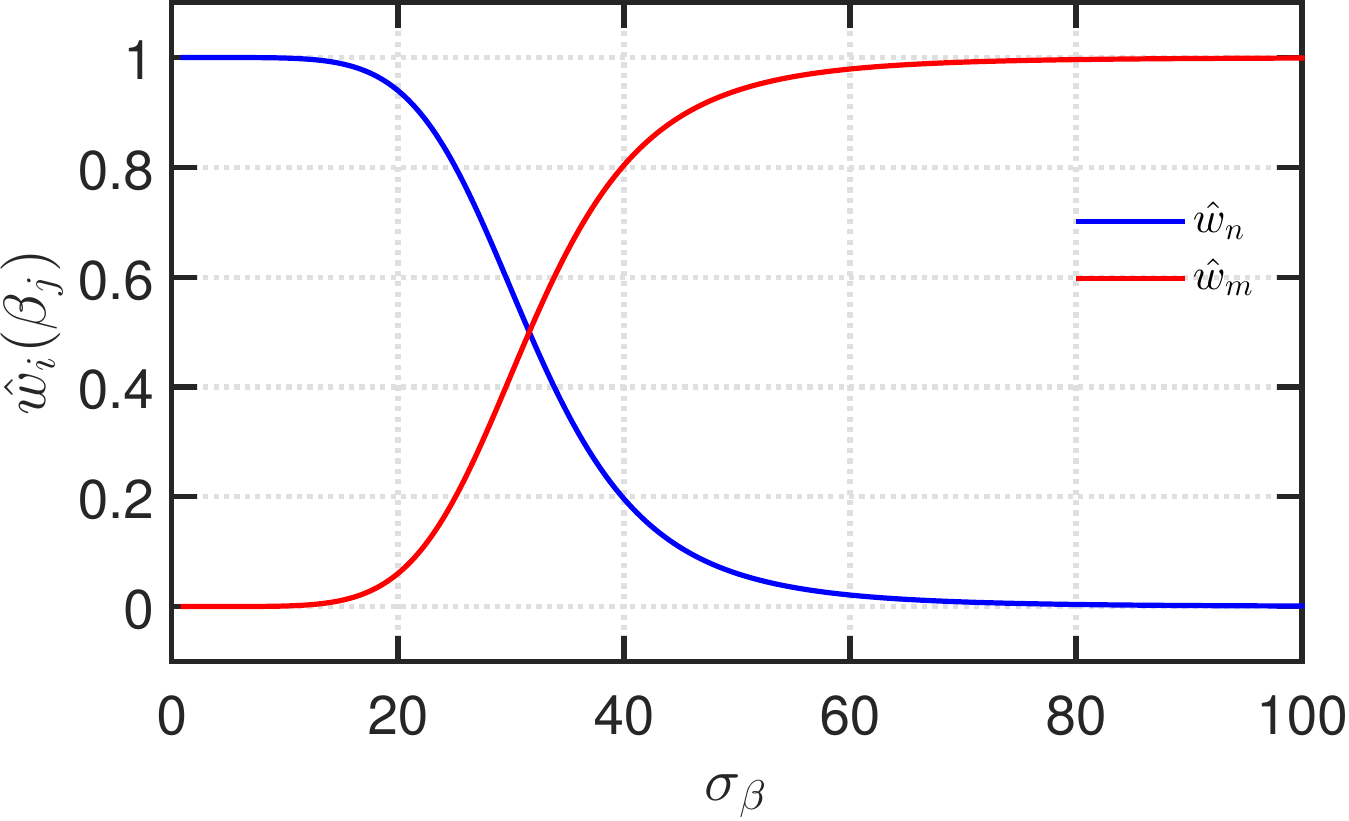}
\end{center}

\caption{The variation of the non-linear weights $\hat{w}_{n}$ and $\hat{w}_{m}$ with the relative smoothness ratio $\sigma_{\beta}$. Larger values $(\sigma_{\beta}\gg 1)$ indicate the underlying data is locally non-smooth.}

\end{figure}

Extending ideas introduced by \citet{dumbser2007arbitrary} and \citet{blossey2008selective}, a heuristic strategy is proposed for the determination of the non-linear weights $\hat{w}_{i}$, based on the distribution of $\beta_{i,j}$ over the stencil $\mathcal{S}_{i} = \{{i-w},\dots,{i+w}\}$, where $w$ is the width of the stencil associated with each grid-cell. A pair of non-linear weights are defined, such that
\begin{eqnarray}
\label{eqn_weight_null}
w_{i,n} = \frac{\lambda_{n}}{\Big(\bar{\epsilon}+\underset{j\in\mathcal{S}_{i}}{\max}\,\left(\beta_{i,j}\right)\Big)^{_r}},
\qquad
\label{eqn_weight_mono}
w_{i,m} = \frac{\lambda_{m}}{\Big(\bar{\epsilon}+\underset{j\in\mathcal{S}_{i}}{\min}\,\left(\beta_{i,j}\right)\Big)^{_r}}
\end{eqnarray}
where $\lambda_{n},\lambda_{m}\in\mathbb{R}^{+}$ are the so-called \textit{linear-weights} associated with the natural and monotone-limited profiles, $r\in\mathbb{Z}^{+}$ is a scalar coefficient that controls the non-linearity of the weighting scheme, and $\bar{\epsilon}$ is a small constant, introduced to prevent division by zero. The normalised non-linear weights are subsequently obtained via a simple re-scaling
\begin{eqnarray}
\label{eqn_weights_nonlin}
\hat{w}_{i,n} = \frac{w_{i,n}}{w_{i,n}+w_{i,m}}\,,\qquad
\hat{w}_{i,m} = \frac{w_{i,m}}{w_{i,n}+w_{i,m}}
\end{eqnarray}
where $\hat{w}_{i,n}+\hat{w}_{i,m}=1$ by construction. Examining expressions (\ref{eqn_weight_null})--(\ref{eqn_weights_nonlin}), it can be seen that the non-linear weights $\hat{w}_{i,n}$ and $\hat{w}_{i,m}$ are determined based on the \textit{relative} difference in $\beta_{i,j}$ over the stencil $\mathcal{S}_{i}$. Specifically, when the ratio
\begin{eqnarray}
\sigma_{\beta}=\frac{\max\,\left(\beta_{i,j}\right)}{\min\,\left(\beta_{i,j}\right)}
\end{eqnarray} 
is sufficiently small, such that $\sigma_{\beta}\simeq 1$, the denominators in expressions (\ref{eqn_weight_null}) are of similar magnitude, and the normalised weights approximate the linear values, such that $\hat{w}_{i,n}\simeq \lambda_{n}$ and $\hat{w}_{i,m}\simeq \lambda_{m}$ as a result. Conversely, when the stencil contains sufficiently different smoothness indicators, such that $\sigma_{\beta}\gg 1$, the relative magnitude of $\hat{w}_{i,m}$ is increased. This effect is amplified by selecting larger values of the exponent $r$, increasing the non-linear response of the scheme. 

Such behaviour can be used to construct a non-linear slope-limiting strategy that combines good accuracy and non-oscillatory characteristics. By selecting a sufficiently large linear weight $\lambda_{n}\gg 1$ for the natural profile and a correspondingly small linear weight $\lambda_{m}\simeq 1$ for the monotone-limited profile, the resulting non-linear weights tend to $\hat{w}_{i,n}\rightarrow 1$ and $\hat{w}_{i,m}\rightarrow 0$ when the underlying data is sufficiently smooth. Such behaviour ensures that the high-order accurate interpolant $Q_{i}(x)\simeq\hat{Q}_{i,n}(x)$ is selected in such cases. Conversely, by selecting a sufficiently large exponent $r > 1$, the non-linear weights tend to $\hat{w}_{i,n}\rightarrow 0$ and $\hat{w}_{i,m}\rightarrow 1$ when the underlying data is non-smooth, ensuring that the robust and non-oscillatory monotone-limited profile  $Q_{i}(x)\simeq\hat{Q}_{i,m}(x)$ is selected. The rate of transition between these two limiting states is governed by the choice of the exponent $r$. Following \citet{dumbser2007arbitrary}, values of $\lambda_{n}={10^{9}}$, $\lambda_{m}={10^{0}}$, $r=6$ and $\bar{\epsilon}={10^{-12}}$ are used throughout this study. See Figure~\ref{figure_weno_weights} for an illustration of the relationship between the non-linear weights and the relative smoothness measure $\sigma_{\beta}$.

In \citet{blossey2008selective}, a related non-oscillatory methodology was used to achieve a so-called \textit{selective monotonicity-preserving} PPM reconstruction, where the natural PPM profiles $\hat{Q}_{i,n}(x)$ were replaced with a monotone reconstruction $\hat{Q}_{i,m}(x)$ when the relative smoothness ratio was sufficiently large ($\sigma_{\beta}\geq \bar{\sigma}_{\beta}$). This smoothness threshold, $\bar{\sigma}_{\beta}$ was specified as a user-defined parameter. While such a scheme was designed to achieve similar outcomes to the WENO-type slope-limiting techniques presented here, the method of `switching' between the natural and monotone-limited profiles represents a key difference in approach. Specifically, in \citet{blossey2008selective}, the polynomial reconstruction is assembled in a discontinuous fashion, with purely unlimited polynomials $Q_{i}(x)\leftarrow\hat{Q}_{i,n}(x)$ selected when $\sigma_{\beta} < \bar{\sigma}_{\beta}$, and purely monotone profiles $Q_{i}(x)\leftarrow\hat{Q}_{i,m}(x)$ selected otherwise. This strategy appears to be similar in spirit to the original Essentially Non-Oscillatory (ENO) methods of \citet{harten1987uniformly}. In contrast, the slope-limiting strategy presented in the current work is based on a continuous blending of the natural and monotone profiles $\hat{Q}_{i,n}(x)$ and $\hat{Q}_{i,m}(x)$, consistent with WENO-type techniques. 

Application of the WENO-type slope-limiting techniques to the PPM and PQM interpolants described in Sections~\ref{section_reconstruction}--\ref{section_monotone_limiter} can be summarised as follows:
\begin{enumerate}
\item Compute and store the oscillation indicator coefficients (\ref{eqn_oscl}) for each grid-cell. Specifically, the cell-centred first- and second-derivatives of the indicator polynomials $\pi_{i}(x)$ are calculated and stored.

\item Compute the natural and monotone-limited grid-cell polynomials $\hat{Q}_{i,n}(x)$ and $\hat{Q}_{i,m}(x)$, as per Sections~\ref{section_reconstruction}--\ref{section_monotone_limiter}.

\item Blend the natural and monotone-limited profiles within each grid-cell according to (\ref{eqn_weno_blend}), using the WENO-like non-linear weights computed from (\ref{eqn_oscl}).
\end{enumerate}

\section{A Semi-Lagrangian formulation for advective transport}
\label{section_lagrangian}

The construction of robust, high-order accurate numerical methods for the solution of advective transport problems remains a critical and challenging problem in many areas of physical modelling. The transport of a scalar quantity $q = q(\mathbf{x},t)$ is subject to
\begin{eqnarray}
\label{eqn_model_transport}
\frac{\partial q}{\partial t} + \nabla\cdot\left(\mathbf{u}q\right) = S,
\end{eqnarray}
where $\mathbf{u} = \mathbf{u}(\mathbf{x},t)$ is a general velocity field and $S = S(\mathbf{x},t)$ represents sources and sinks of the quantity $q$. In this section, the high-order piecewise polynomial interpolants described in Sections~\ref{section_reconstruction}--\ref{section_weno_limiter}, are used to define a semi-Lagrangian framework for the solution of (\ref{eqn_model_transport}). The construction of both one-dimensional and direction-split multi-dimensional schemes are discussed.

\subsection{A one-dimensional framework}

\medskip

Assuming a source-free, one-dimensional flow, the advective transport equation (\ref{eqn_model_transport}) can be written in a Lagragian frame of reference, such that
\begin{eqnarray}
\label{eqn_transport_lagrangian}
\frac{D q}{D t} = 0,
\end{eqnarray}
where $\nicefrac{D(\cdot)}{D t} = \nicefrac{\partial(\cdot)}{\partial t} + \nabla\cdot(\mathbf{u}(\cdot))$ is the material derivative. Adopting a finite-volume discretisation of (\ref{eqn_transport_lagrangian}) over a set of deforming control-volumes $\Omega(x,t)$ that move with the velocity $u(x,t)$, the discrete variables are updated at each time-step according to
\begin{eqnarray}
\label{eqn_transport_integral}
\left(\int_{\Omega_{i}} Q(x)\,\mathrm{d}x \right)_{i}^{t+\Delta t} = 
\left(\int_{\Omega_{i}} Q(x)\,\mathrm{d}x \right)_{i}^{t},
\end{eqnarray}
where (\ref{eqn_transport_integral}) is simply a statement of exact mass conservation. Compared to flux-based formulations, it is important to note that the use of (\ref{eqn_transport_integral}) does not impose restrictions on the size of the time-step due to CFL-type constraints \citep{courant1967partial}. The discretisation of (\ref{eqn_transport_lagrangian}) is completed by selecting a suitable time-integration scheme for the evolution of the control-volumes $\Omega(x,t) = x_{i+1}-x_{i}$, here, simply the one-dimensional grid-cells spanning the points $x_{i}$. In the semi-Lagrangian framework, the position of the points $x_{i}^{n+1}$ at $t = t^{n+1}$ are simply the positions of the fixed one-dimensional target grid. Correspondingly, the position of the so-called \textit{departure-points} $x_{i}^{*}$ at $t=t^{n}$ can be found by integrating the set of \textsc{ode}'s 
\begin{eqnarray}
\label{eqn_departure}
\frac{d x_{i}^{*}}{d t} = u(x,t),\quad x_{i}^{*}(t^{n+1}) = x_{i}
\end{eqnarray}
backwards in time, from $t = t^{n+1}$ to $t = t^{n}$. Such a process positions the points $x_{i}^{*}$ according to the \textit{characteristics} associated with the flow. In this study, (\ref{eqn_departure}) is integrated using a fourth-order Runge-Kutta method \citep{butcher1996history}. Once the positions of the departure points have been calculated, the integral term on the right hand side of (\ref{eqn_transport_integral}) is evaluated via a two-step process. Firstly, a piecewise polynomial reconstruction $Q_{i}^{t}(x)$ is computed using the existing cell-mean data $\bar{q}_{i}^{n}$ on the current mesh $x_{i}^{n}$. Secondly, the grid-cell integral terms appearing on the right hand side of (\ref{eqn_transport_integral}) are evaluated by integrating $Q_{i}^{n}(x)$ over the deformed departure grid $x_{i}^{*}$, via
\begin{eqnarray}
\label{eqn_lagrangian_cell_ints}
\left(\int_{\Omega_{i}} Q(x)\,\mathrm{d}x \right)_{i}^{t} = \sum_{j} \int_{x_{j}^{n}}^{x_{j+1}^{n}}Q_{j}^{n}(x)\,\mathrm{d}x,\quad \forall \Omega_{j}^{n}\cap\Omega_{i}^{*}\neq\emptyset,
\end{eqnarray}
where the summation is taken over the set of all grid-cells $j$ in the existing mesh $\Omega_{j}^{n}$ that intersect with a departure cell $\Omega_{i}^{*}$. The time-stepping procedure is completed by computing the new cell-mean distribution from (\ref{eqn_lagrangian_cell_ints}), such that
\begin{eqnarray}
\bar{q}_{i}^{n+1} = \frac{1}{h_{i}} \left(\sum_{j} \int_{x_{j}^{n}}^{x_{j+1}^{n}}Q_{j}^{n}(x)\,\mathrm{d}x\right).
\end{eqnarray}

A single step of the one-dimensional semi-Lagrangian algorithm described previously can be summarised as follows:
\begin{enumerate}
\item Calculate the position of the departure points $x_{i}^{*}$ by integrating (\ref{eqn_departure}) backwards in time, from $t=t^{n+1}$ to $t = t^{n}$.

\item Reconstruct the piecewise polynomial interpolants $Q_{i}^{t}(x)$ from the cell-mean data $\bar{q}_{i}^{n}$ on the existing mesh $x_{i}^{n}$ at $t=t^{n}$, using either the PPM or PQM reconstructions described in Sections~\ref{section_reconstruction}, \ref{section_monotone_limiter} and  \ref{section_weno_limiter}.

\item Compute the grid-cell integrals (\ref{eqn_lagrangian_cell_ints}), by integrating the polynomial profiles $Q_{i}^{n}(x)$ over the deformed departure control volumes $\Omega_{i}^{*} = x_{i+1}^{*}-x_{i}^{*}$ via (\ref{eqn_lagrangian_cell_ints}).

\item Calculate the new cell-mean distribution $\bar{q}_{i}^{n+1}$ by scaling the grid cell integrals (\ref{eqn_lagrangian_cell_ints}) by the grid-cell widths $h_{i}$.
\end{enumerate} 

\subsection{Multiple dimensions: direction splitting}

\medskip

The one-dimensional semi-Lagrangian algorithm described previously can be extended to handle multi-dimensional advection problems using a Strang splitting approach \citep{easter1993two,blossey2008selective}
\begin{eqnarray}
\bar{q}_{i,j}^{(1)} = \frac{1}{\Delta x_{i}}\left(\sum_{k} \int_{x_{k}^{(n)}}^{x_{k+1}^{(n)}}Q_{k}^{(n)}(x)\,\mathrm{d}x\right)
\\
\bar{q}_{i,j}^{n+1} = \frac{1}{\Delta y_{j}}\left(\sum_{k} \int_{y_{j}^{(1)}}^{y_{k+1}^{(1)}}Q_{k}^{(1)}(y)\,\mathrm{d}y\right)
\end{eqnarray}
where each sub-step is an application of the full one-dimensional algorithm described previously. A second-order accurate time integration is achieved by exchanging the order of the $x$ and $y$ integrations at odd or even numbered time-steps, respectively.

\section{Experimental results}
\label{section_results}

The PPM and PQM piecewise polynomial interpolants presented in Sections~\ref{section_reconstruction}, \ref{section_monotone_limiter}, \ref{section_weno_limiter} and the high-order semi-Lagrangian advection scheme presented in Section~\ref{section_lagrangian}, were applied to a range of benchmark problems designed to test their effectiveness and computational efficiency. The accuracy and relative efficiency of the various monotonicity-preserving and WENO-type slope-limiting strategies was examined in detail. The reconstruction methods were implemented in the PPR (Piecewise Polynomial Reconstruction) library \citep{engwirda2015} using Fortran-95.

\subsection{One-dimensional remapping}

\medskip

The performance of the PPM and PQM reconstructions was assessed using a series of comparative `remapping'-type tests adapted from \citet{white2008high}. These test cases serve as a proxy for the remapping-type operations performed in generalised vertical coordinate atmospheric and oceanic circulation models. Given an initial profile $Q^{(0)}(x)$ and a uniform one-dimensional grid $x_{i}^{(0)}$, a variant of the semi-Lagrangian algorithm described in Section~\ref{section_lagrangian} was used to iteratively `remap' the profile onto a sequence of non-uniform grids. Following \citet{white2008high}, each remapping cycle was implemented in two-steps: first transferring the profile $Q^{n}(x)$ from the uniform grid $x^{(0)}$ to a randomised, non-uniform grid $x^{(1)}$, before reversing the process, and transferring $Q^{(1)}(x)$ back to the uniform grid $x^{(0)}$. In addition to a non-uniform grid-spacing, the intermediate grid $x^{(1)}$ was constructed to contain 10\% fewer grid-points. 

\begin{figure}[t]

\label{figure_1d_smooth}

\begin{center}
\includegraphics[width=.425\textwidth]{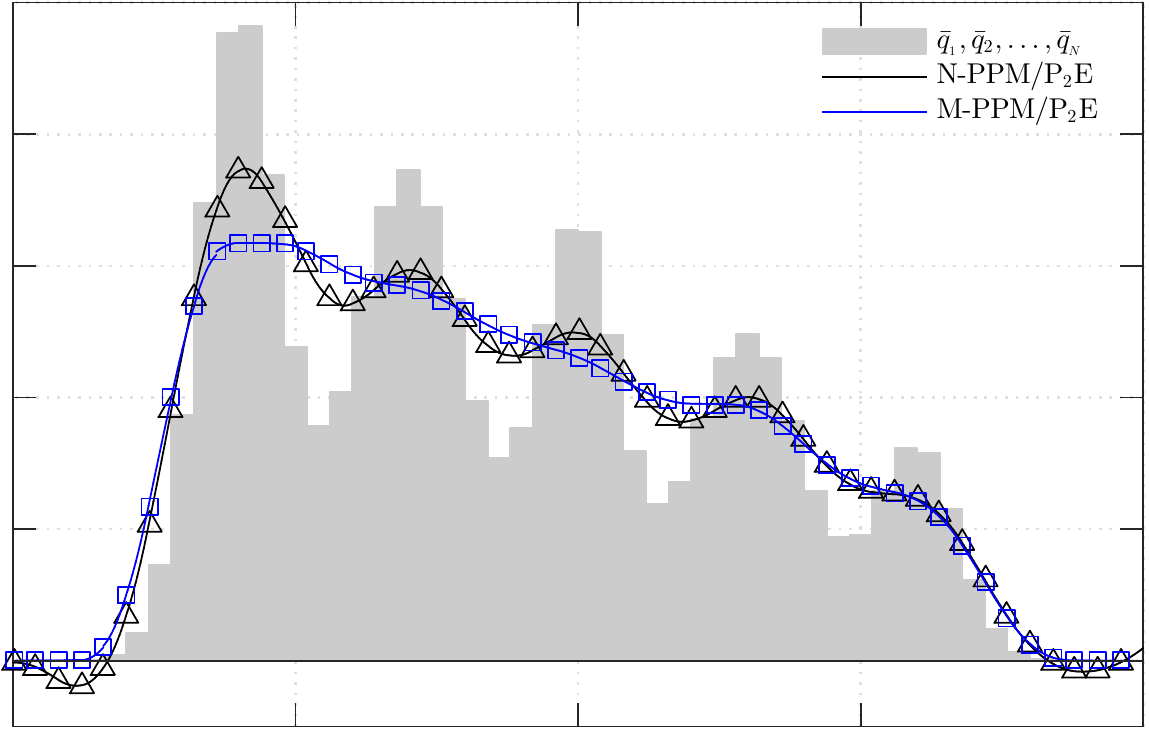}\quad
\includegraphics[width=.425\textwidth]{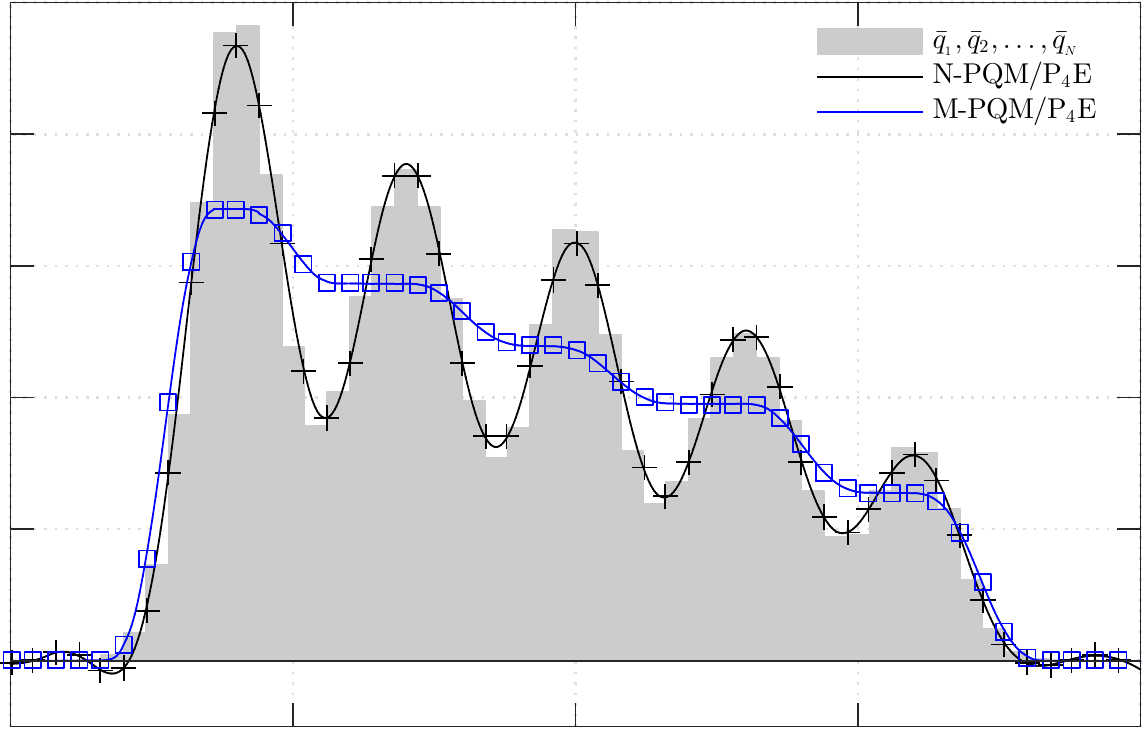}
\\[2.5ex]
\includegraphics[width=.425\textwidth]{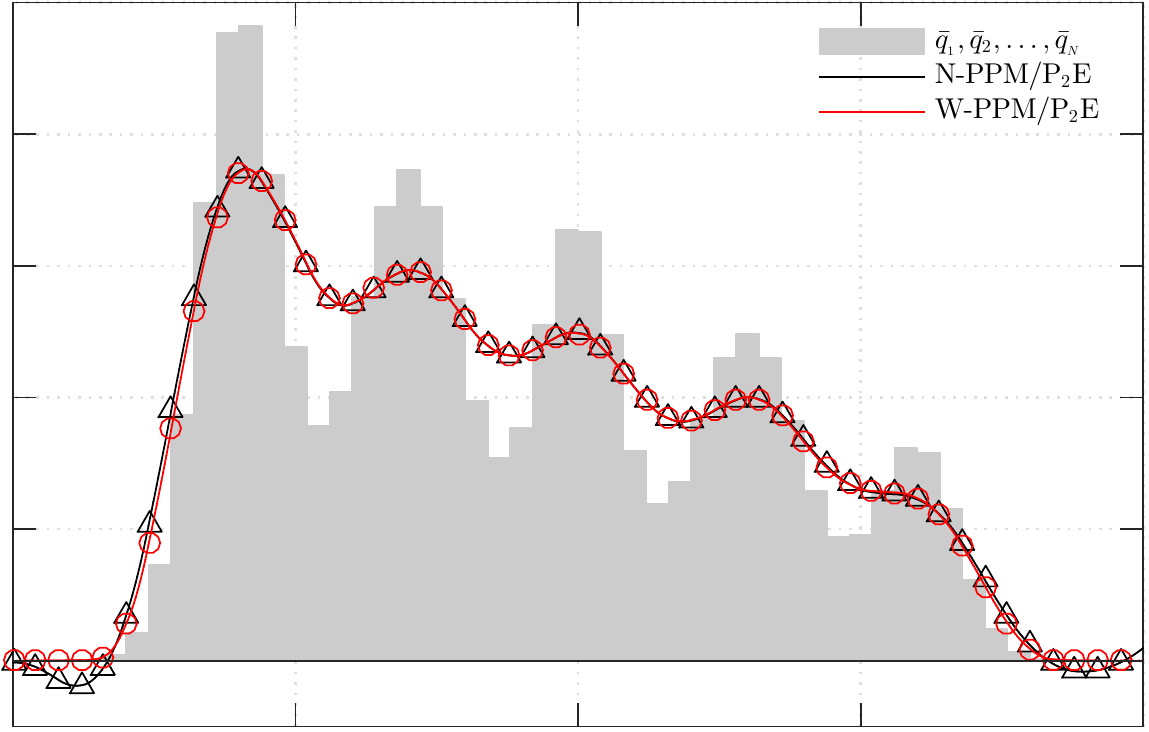}\quad
\includegraphics[width=.425\textwidth]{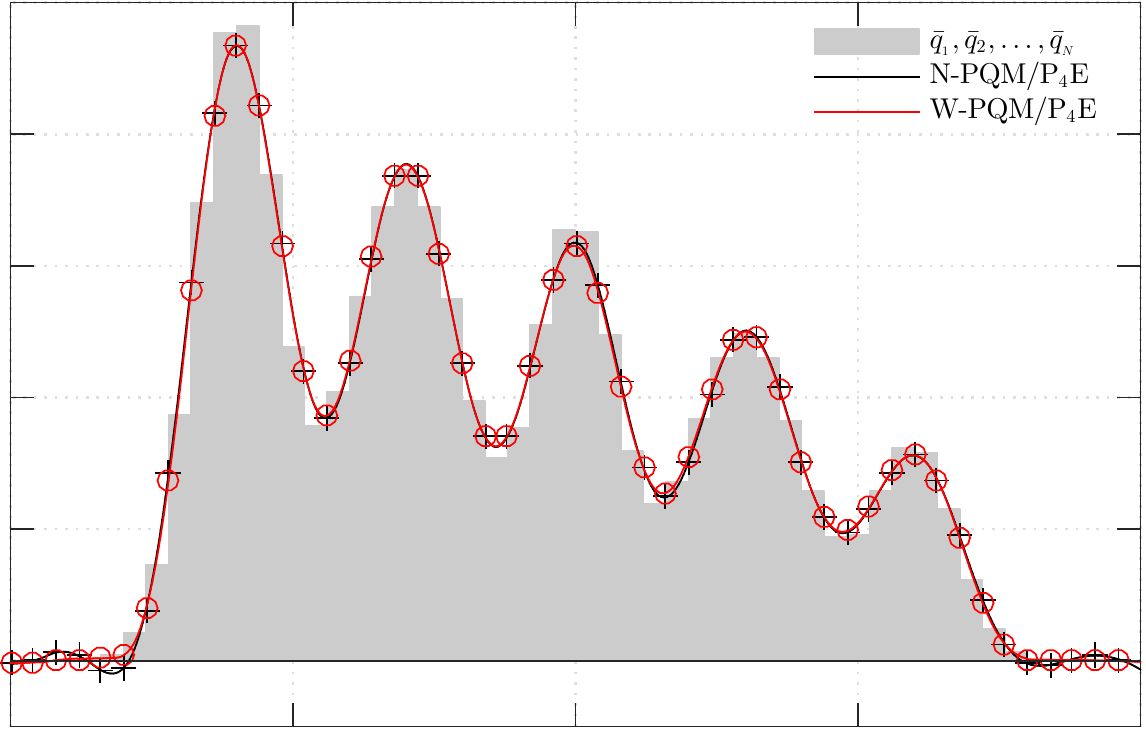}
\end{center}

\caption{Comparison of PPM and PQM reconstruction techniques for the smooth one-dimensional remapping test case at $\mathrm{N} = 50$. Profiles are shown after 250 remapping steps. Series prefixed with an `N-' denote results obtained using the natural (unlimited) reconstructions. Series prefixed with `W-' denote results obtained using the WENO-type limiter. Series prefixed with `M-' denote results obtained using the standard monotone limiters.}

\end{figure}

The performance of the various polynomial reconstructions was first assessed using the smooth initial profile
\begin{eqnarray}
Q^{(0)} =  \operatorname{e}^{-(x + 6)^2} +
\frac{3}{4}\operatorname{e}^{-\frac{1}{2}(x + 3)^2} +
\frac{2}{3}\operatorname{e}^{-x^2} +
\frac{1}{2}\operatorname{e}^{-\frac{1}{2}(x - 3)^2} +
\frac{1}{3}\operatorname{e}^{-(x - 6)^2},
\end{eqnarray}
chosen to provide a simple and yet non-trivial profile for which asymptotic rates of convergence can be expected. The results of the iterative remapping experiments after 250 iterations are shown in Figure~\ref{figure_1d_smooth}, including results for both the PPM and PQM interpolants coupled with either the monotone or WENO-type  slope limiters. Methods prefixed `N-' denote unlimited methods, those prefixed `M-' denote the monotone schemes, and those prefixed `W-' refer to the WENO-based techniques. Calculations were performed using $N=50$. Results are shown for the third- and fifth-order accurate P$_2$E and P$_4$E edge estimate schemes, respectively, although performance using the P$_3$E and P$_5$E  approximations was found to be little different. Based on visual inspection, it is clear, firstly, that both the monotone and WENO limited PQM interpolants offer improved accuracy compared to the equivalent PPM schemes. Recalling that the PQM interpolant is a nominally fifth-order accurate method in contrast to the third-order accurate PPM scheme, such results are consistent with expectations. Secondly, it is clear that the use of the WENO-type slope-limiters for both the PPM and PQM reconstructions leads to significantly improved accuracy, with both methods resolving the smooth peaks present in the profile. In the case of the W-PQM schemes the level of induced numerical dissipation is so low that the reconstructed profile is essentially indistinguishable from the exact function. In contrast, both the M-PPM and M-PQM schemes result in a clear flattening of the smooth peaks present in the underlying profile. Noting that the W-PPM and W-PQM schemes are no more diffusive than the unlimited N-PPM and N-PQM reconstructions, it is clear that the WENO-type slope-limiter imposes very low levels of additional numerical dissipation.

\begin{figure}[t]

\label{figure_error_1d}

\begin{center}
\includegraphics[width=.65\textwidth]{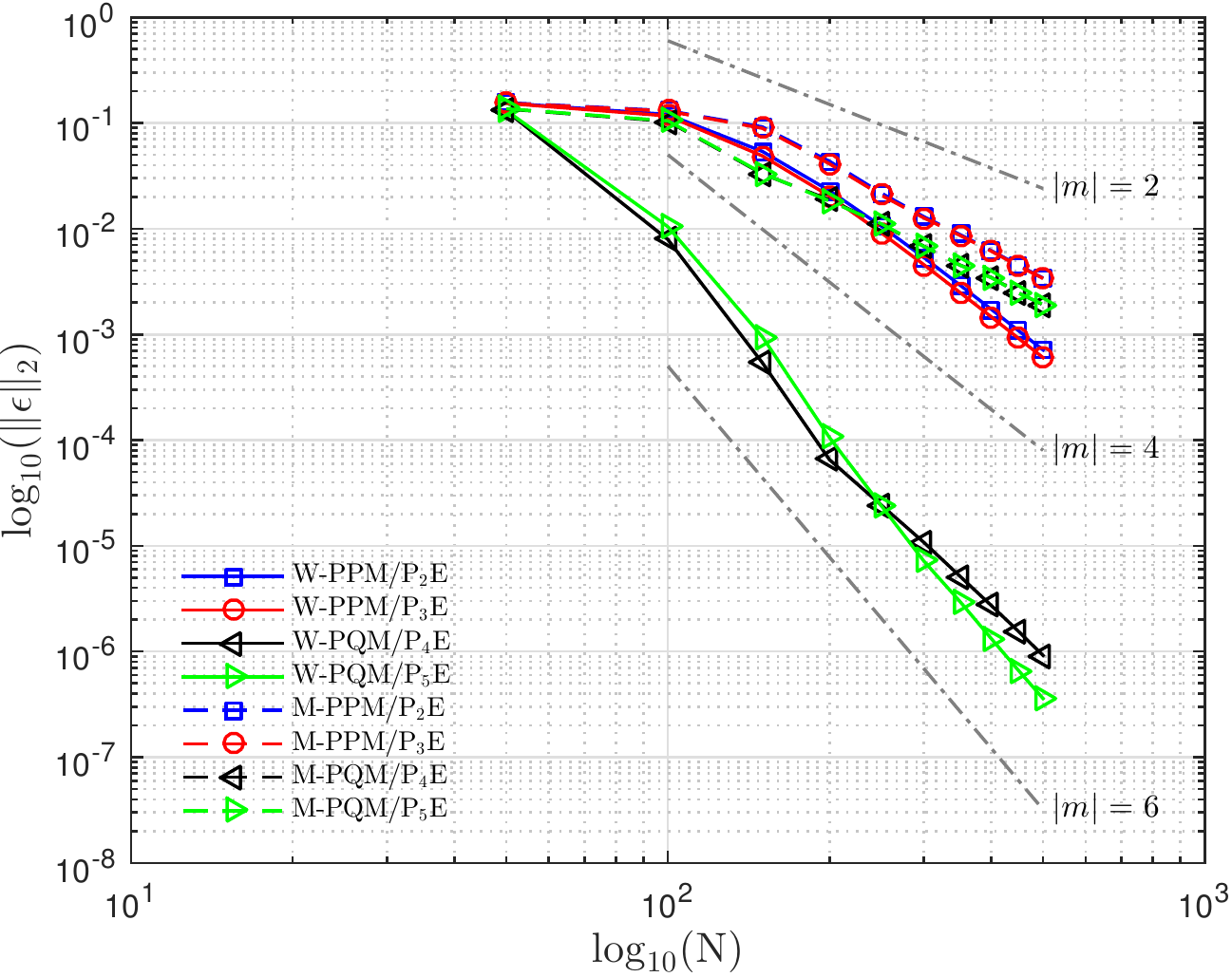}
\end{center}

\caption{Results for the one-dimensional remapping test case, showing the convergence of $L_{2}$-error ($\|\epsilon\|_{2}$) with the number of grid-cells ($\mathrm{N}$). Errors are calculated after 10,000 remapping steps. Series prefixed with `W-' denote results obtained using the WENO-type limiter. Series prefixed with `M-' denote results obtained using the standard monotone limiters. Additional series illustrating $|m|=\{2,4,6\}$-th order convergence are shown for convenience.}

\end{figure}

The performance of the various interpolants was assessed quantitatively, with a mesh refinement study used to establish both the effective order of accuracy of each of the methods, and to contrast their relative computational efficiencies. In Figure~\ref{figure_error_1d}, the results of the mesh refinement study are presented, showing the variation in the $L_{2}$-errors with increasing mesh resolution. Results were reported after 10,000 remapping cycles, and were obtained for the various combinations of interpolant, slope-limiter and edge approximation schemes available. Most importantly, these results illustrate the effectiveness of the new WENO-type slope-limiter, showing that both the PPM and PQM based methods achieve fully third- and fifth-order accuracy when the WENO-type limiter is selected. In contrast, methods based on the monotone schemes are seen to display only second-order behaviour, irrespective of the nominal order of accuracy of the underlying PPM or PQM reconstructions. In addition to improved asymptotic performance, the WENO limited schemes are also seen to outperform the monotone methods in terms of absolute error magnitudes, with the W-PQM scheme reducing the $L_{2}$-error by up to five orders of magnitude when compared to M-PQM. The W-PPM scheme was typically found to outperform both the M-PQM and M-PPM schemes by a smaller margin. Results using the various cell- and edge-centred approximation schemes showed little variation.

\begin{figure}[t]
\label{figure_times_1d}

\begin{center}
\includegraphics[width=.65\textwidth]{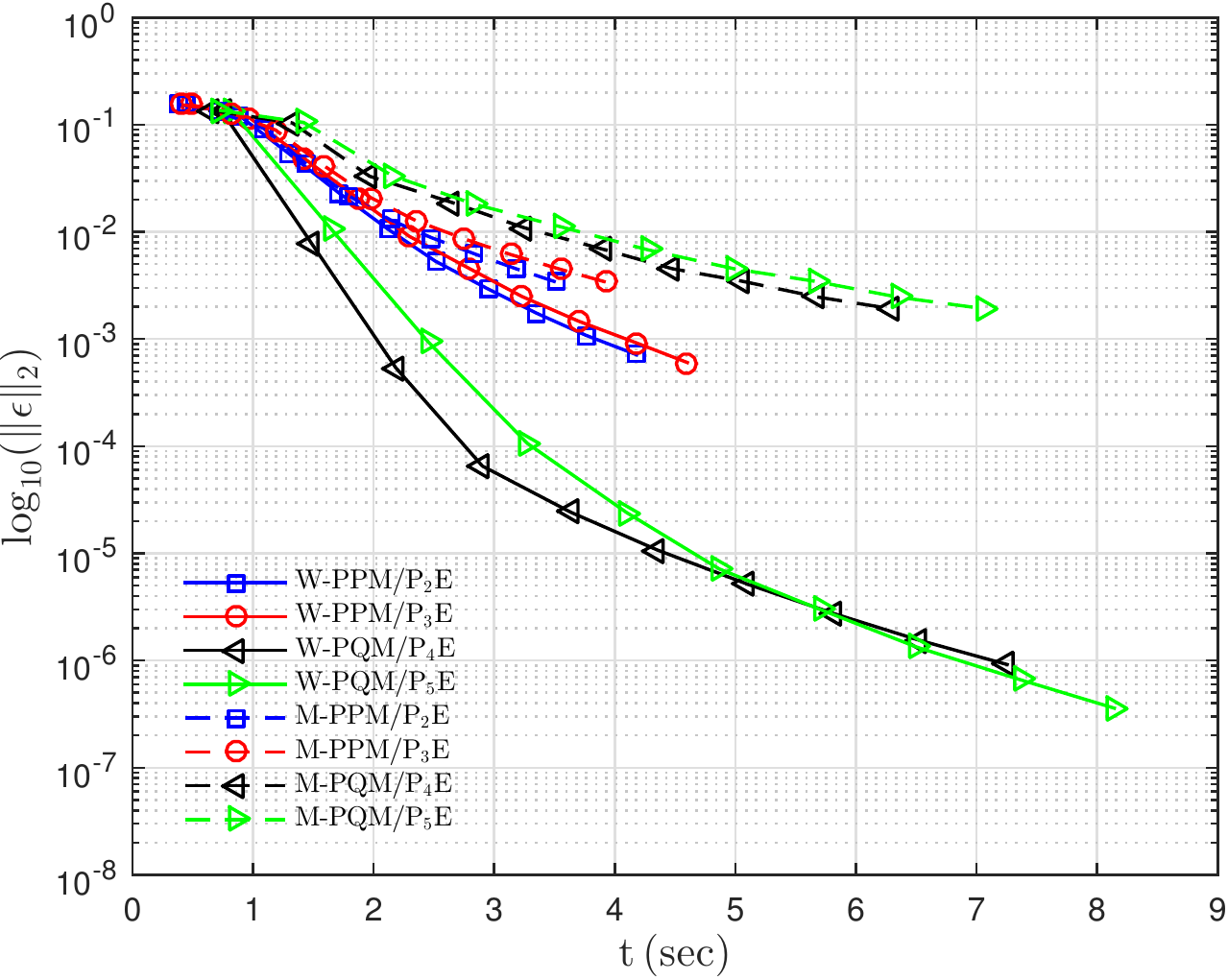}
\end{center}

\caption{Results for the one-dimensional remapping test case, showing the relationship between $L_{2}$-error ($\|\epsilon\|_{2}$) and overall runtime ($\mathrm{t}$). Tests were run using a 2.4GHz Intel i7 processor. Errors are calculated after 10,000 remapping steps. Times are reported in seconds and are the average of three runs. Series prefixed with `W-' denote results obtained using the WENO-type limiter. Series prefixed with `M-' denote results obtained using the standard monotone limiters. Results biased towards the lower-left portion of the graph indicate greater relative efficiency.}

\end{figure}

The relative computational efficiency of the various one-dimensional reconstruction schemes is illustrated in Figure~\ref{figure_times_1d}, showing the total computational effort required to achieve a certain level of $L_{2}$-error for each of the various interpolants. These results again illustrate the benefits of the new WENO-based slope-limiting techniques, showing that both the W-PPM and W-PQM schemes are significantly more efficient than their monotone counterparts. These results also show that when comparing the monotone PPM and PQM interpolants only, there is no net benefit to be gained by using the higher-order PQM scheme, with the gains in accuracy offset by an increased computational burden. Overall, the W-PQM scheme was found to clearly outperform the other candidate schemes, offering significantly reduced error magnitudes at similar levels of computational burden for all but the coarsest resolutions. Considering total computational effort alone, it can be seen that use of the WENO-based schemes W-PPM and W-PQM, result in only marginal increases to overall runtime, with 10--20\% increases observed compared to the respective monotone methods.

\subsection{A one-dimensional example containing sharp features}

\medskip

\begin{figure}[t]

\label{figure_1d_nonsmooth}

\begin{center}
\includegraphics[width=.425\textwidth]{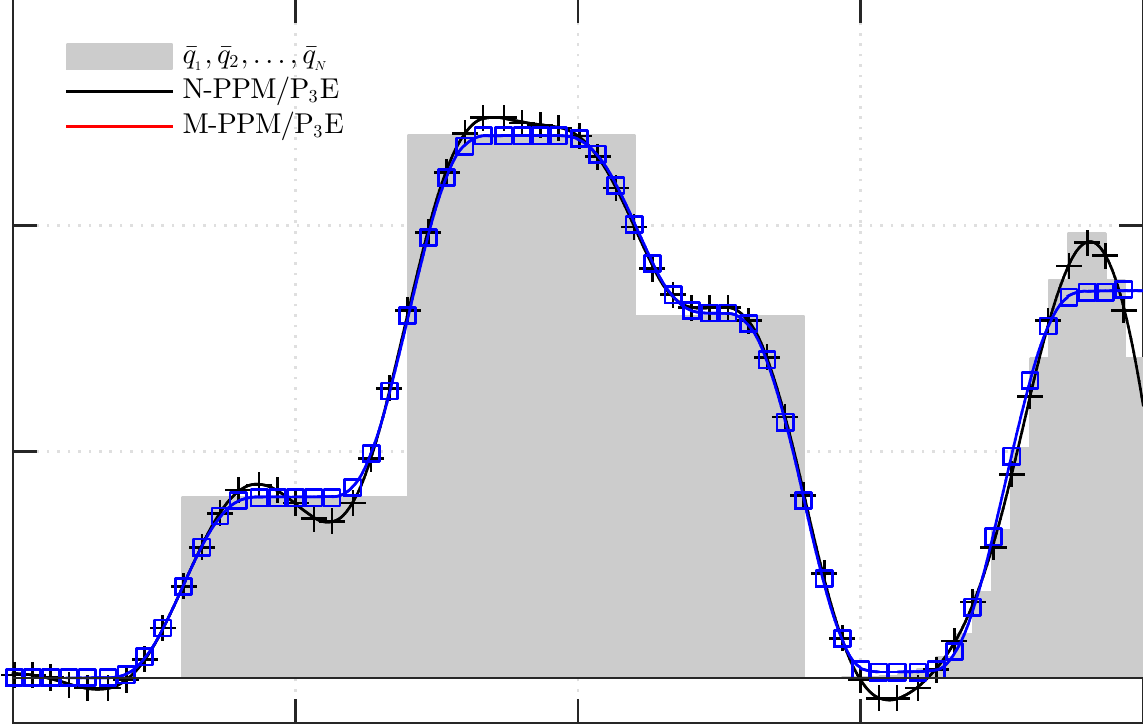}\quad
\includegraphics[width=.425\textwidth]{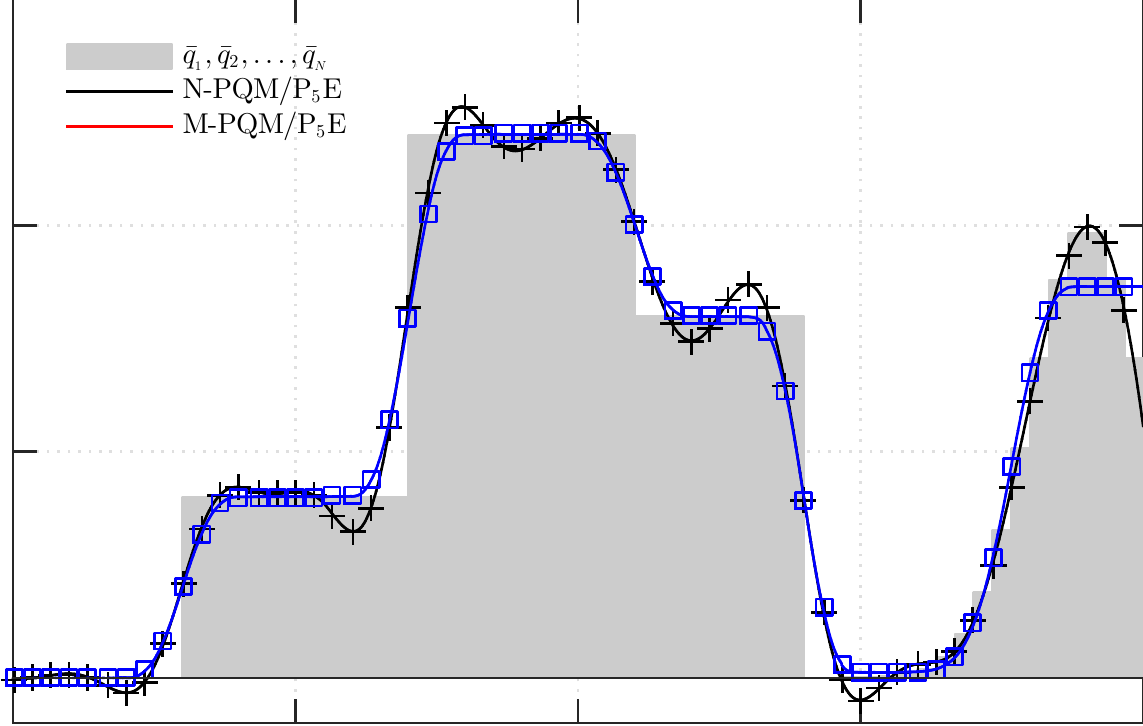}
\\[2.5ex]
\includegraphics[width=.425\textwidth]{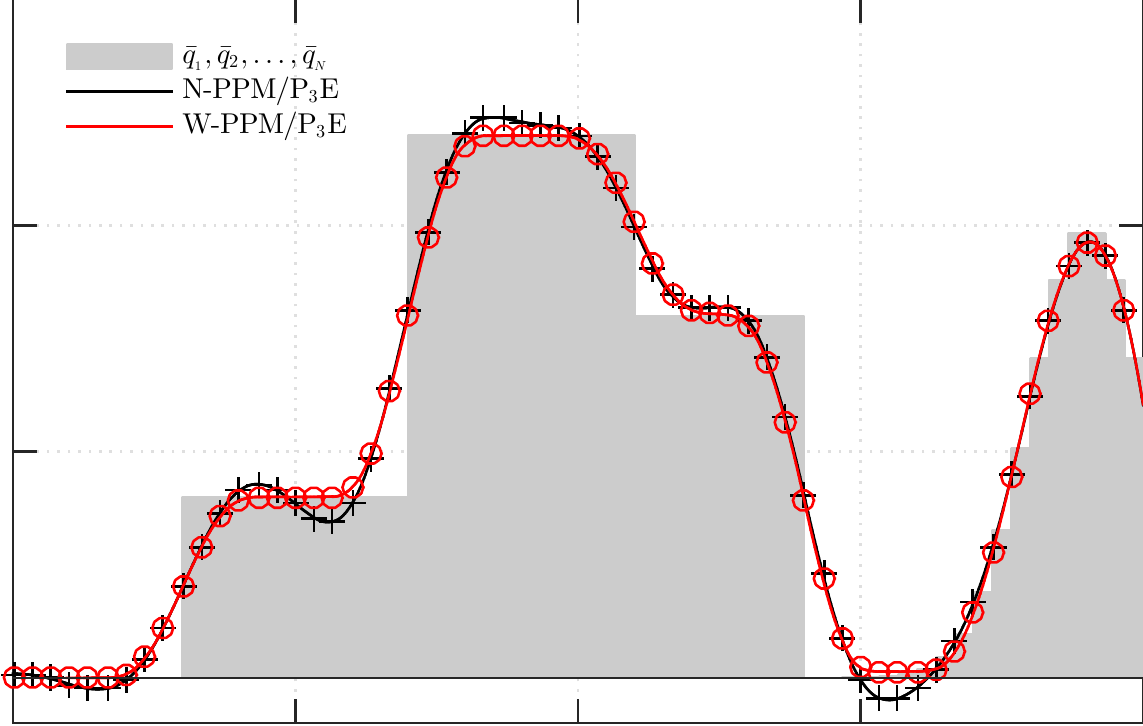}\quad
\includegraphics[width=.425\textwidth]{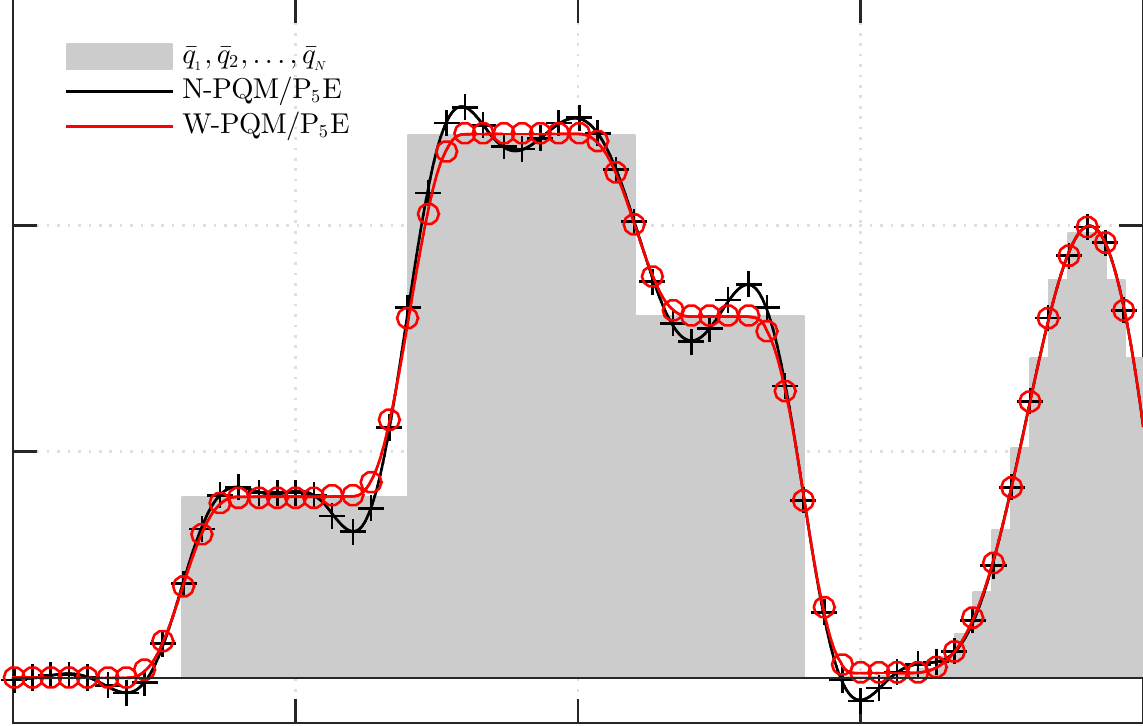}
\end{center}

\caption{Comparison of PPM and PQM reconstruction techniques for the non-smooth one-dimensional remapping test case at $\mathrm{N} = 60$. Profiles are shown after 250 remapping steps. Series prefixed with an `N-' denote results obtained using the natural (unlimited) reconstructions. Series prefixed with `W-' denote results obtained using the WENO-type limiter. Series prefixed with `M-' denote results obtained using the standard monotone limiters.}

\end{figure}

The performance of the various PPM and PQM reconstruction schemes was also assessed for the non-smooth profile
\begin{eqnarray}
\label{eqn_nonsmooth_1d}
Q^{(0)} = \left\{
\begin{array}{cc}
\frac{4}{10}, & \text{if } \left(x \geq -7 \text{ and } x < -3\right) \\[1ex]
\frac{12}{10}, & \text{if } \left(x \geq -3 \text{ and } x < +1\right) \\[1ex]
\frac{8}{10}, & \text{if } \left(x \geq +1 \text{ and } x < +4\right) \\[1ex]
\operatorname{e}^{-\frac{1}{2}(x - 9)^2}, & \text{otherwise}
\end{array}
\right.
\end{eqnarray}
The smooth Gaussian profile in (\ref{eqn_nonsmooth_1d}) is positioned to intersect with the right boundary to test the performance of the boundary extrapolation techniques described in Section~\ref{section_edge_estimates}. Results for both the PPM and PQM interpolants are reported in Figure~\ref{figure_1d_nonsmooth}, showing the various reconstruction profiles after 250 remapping iterations, consistent with previous experiments. In all cases, it is clear that both the slope-limiting and boundary extrapolation strategies perform as expected, with the monotone- and WENO-limited PPM and PQM profiles shown to smoothly interpolate the discontinuous region of the data without overshoots, while the WENO-limited profiles are shown to smoothly extrapolate the Gaussian profile at the right-hand boundary. It is clear that the unlimited N-PPM and N-PQM profiles contain spurious oscillations adjacent to the discontinuous features in the data. Consistent with previous results, it can be seen that WENO-based W-PPM and W-PQM interpolants outperform the respective monotone schemes near the smooth features in the data, with the additional numerical dissipation leading to a flattening of both the M-PPM and M-PQM profiles at the right-hand boundary. Overall, the high-order W-PQM interpolant is again shown to offer superior performance, offering the best representation of both the smooth and discontinuous features in the underlying data.

\subsection{Two-dimensional solid-body rotation}

\medskip

\begin{figure}[t]

\label{figure_solid_rotation}

\begin{center}
\includegraphics[width=.275\textwidth]{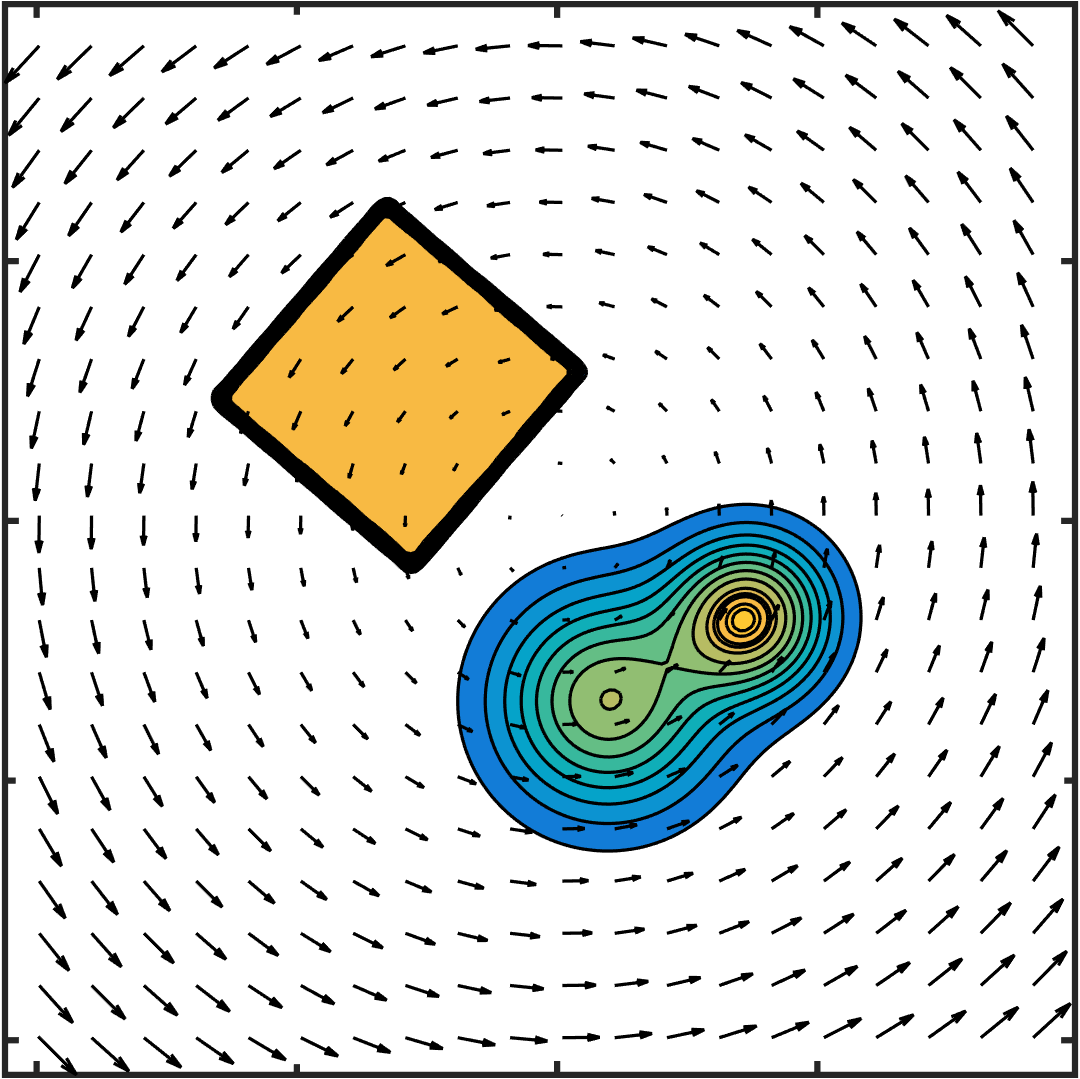}\quad
\includegraphics[width=.275\textwidth]{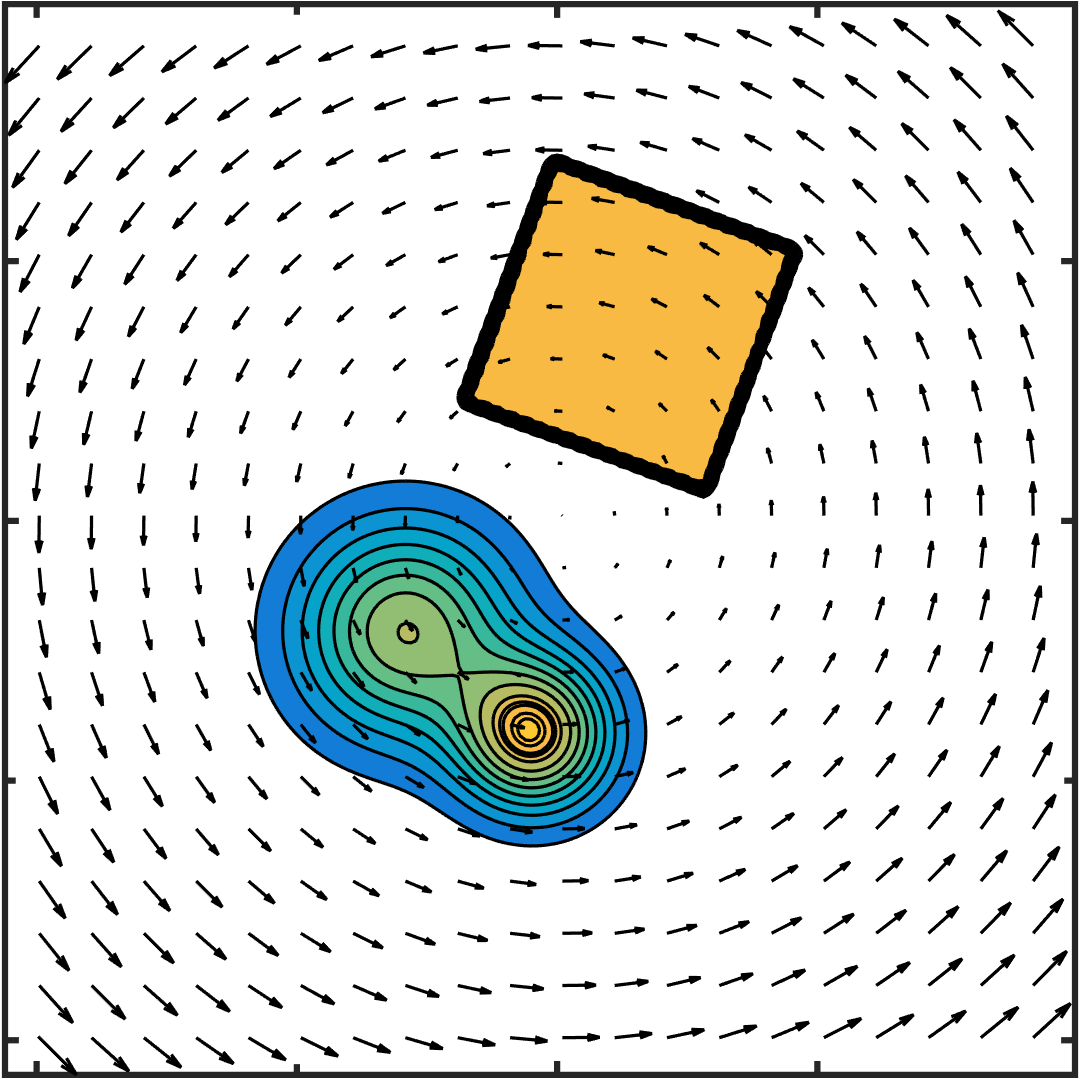}\quad
\includegraphics[width=.275\textwidth]{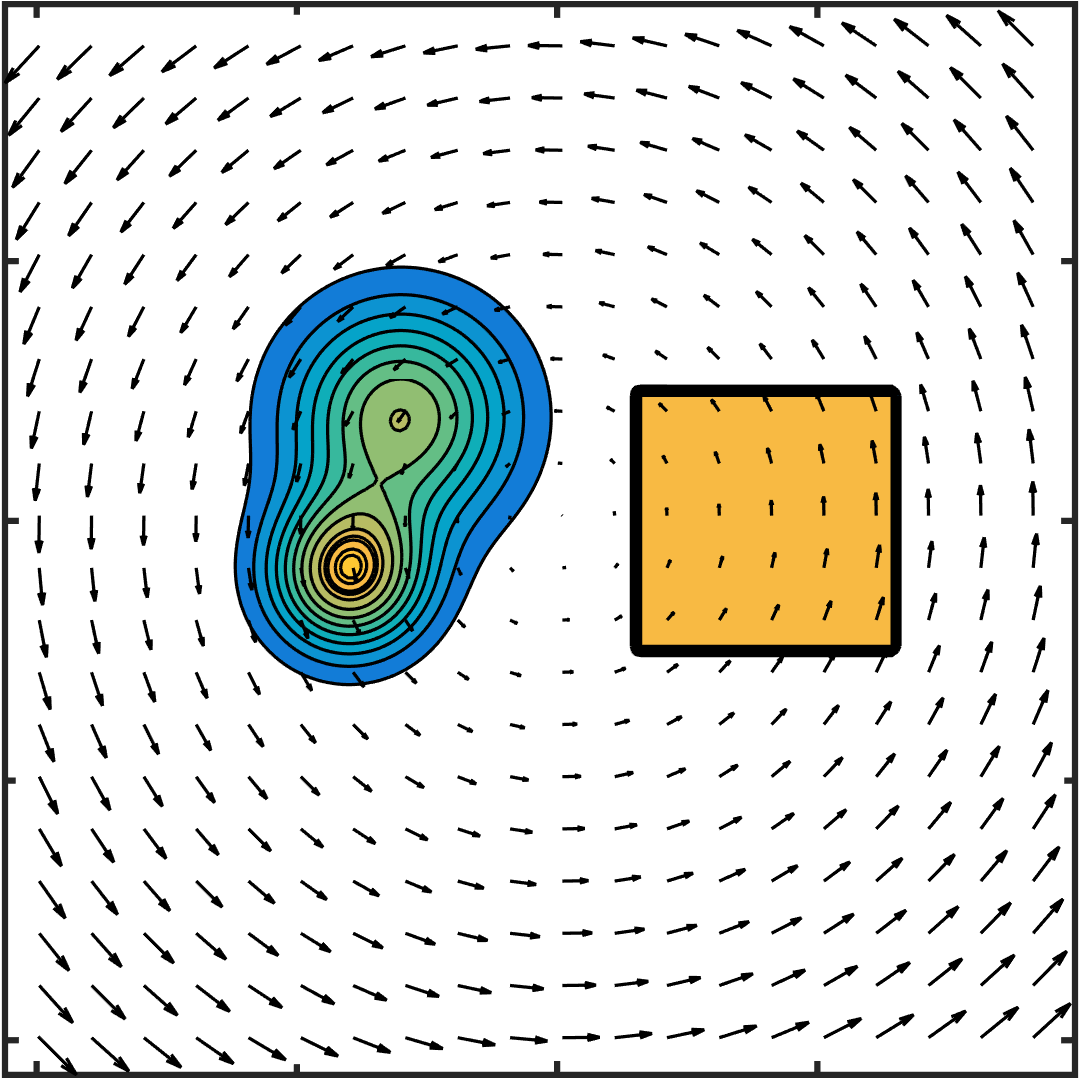}\\[2.25ex]
\includegraphics[width=.275\textwidth]{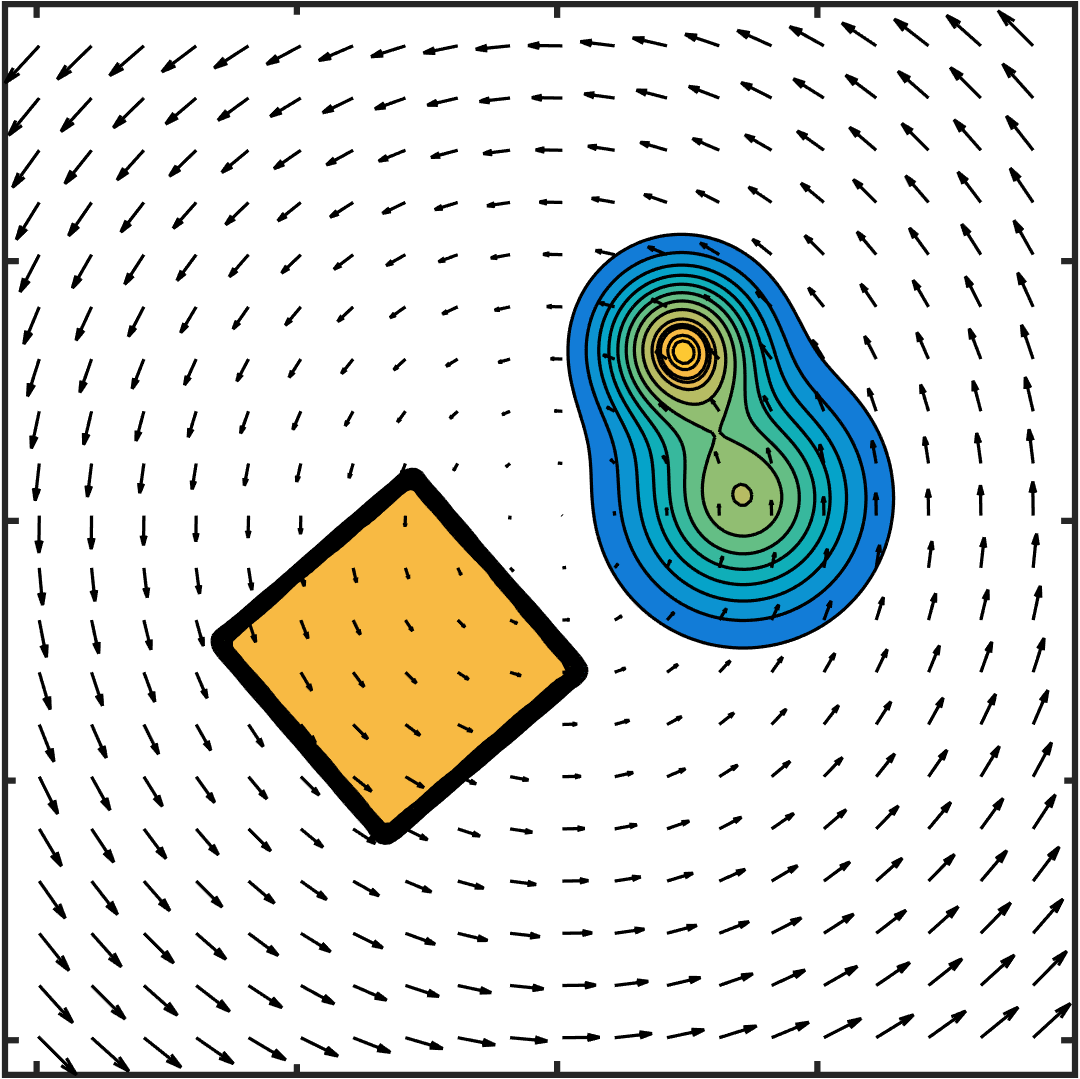}\quad
\includegraphics[width=.275\textwidth]{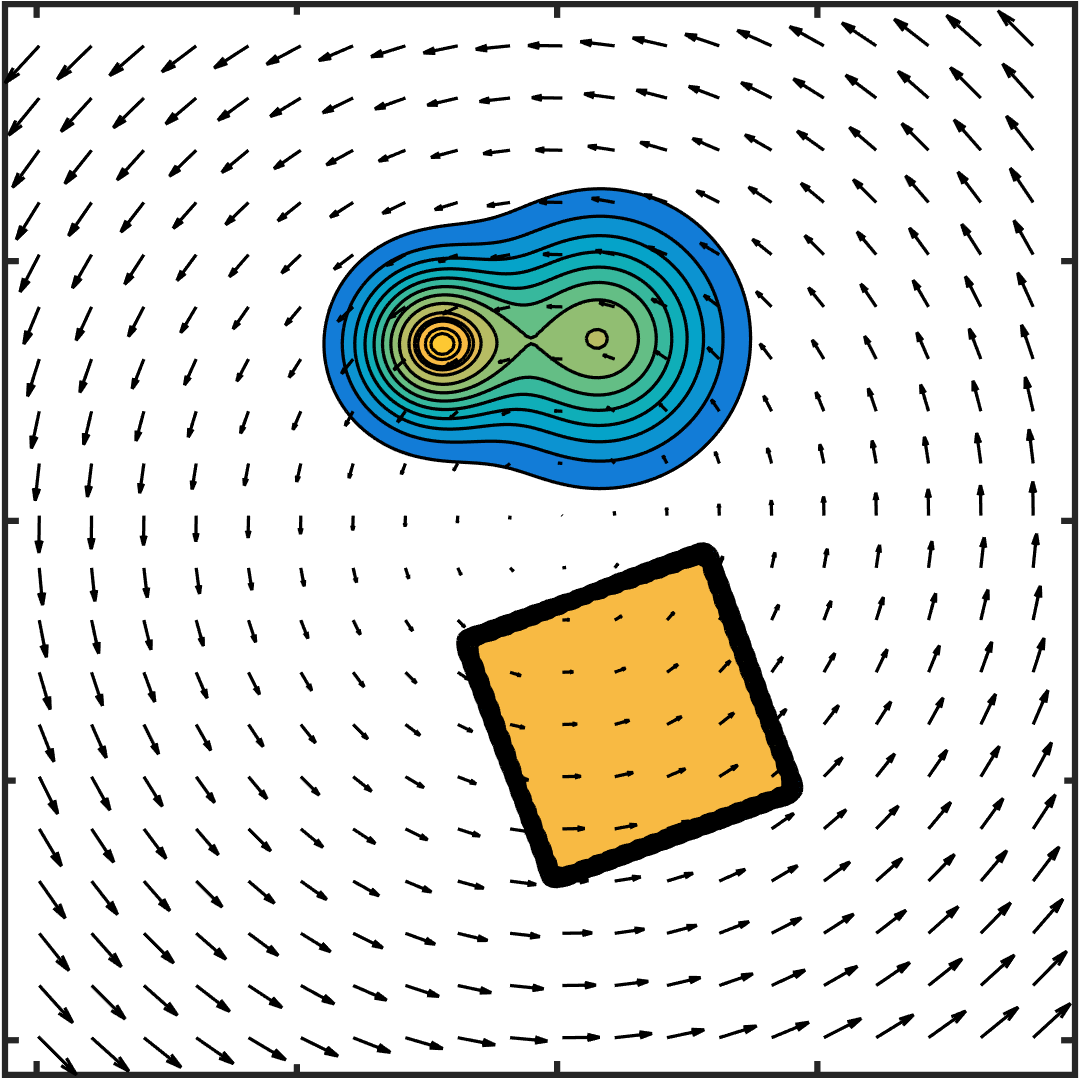}\quad
\includegraphics[width=.275\textwidth]{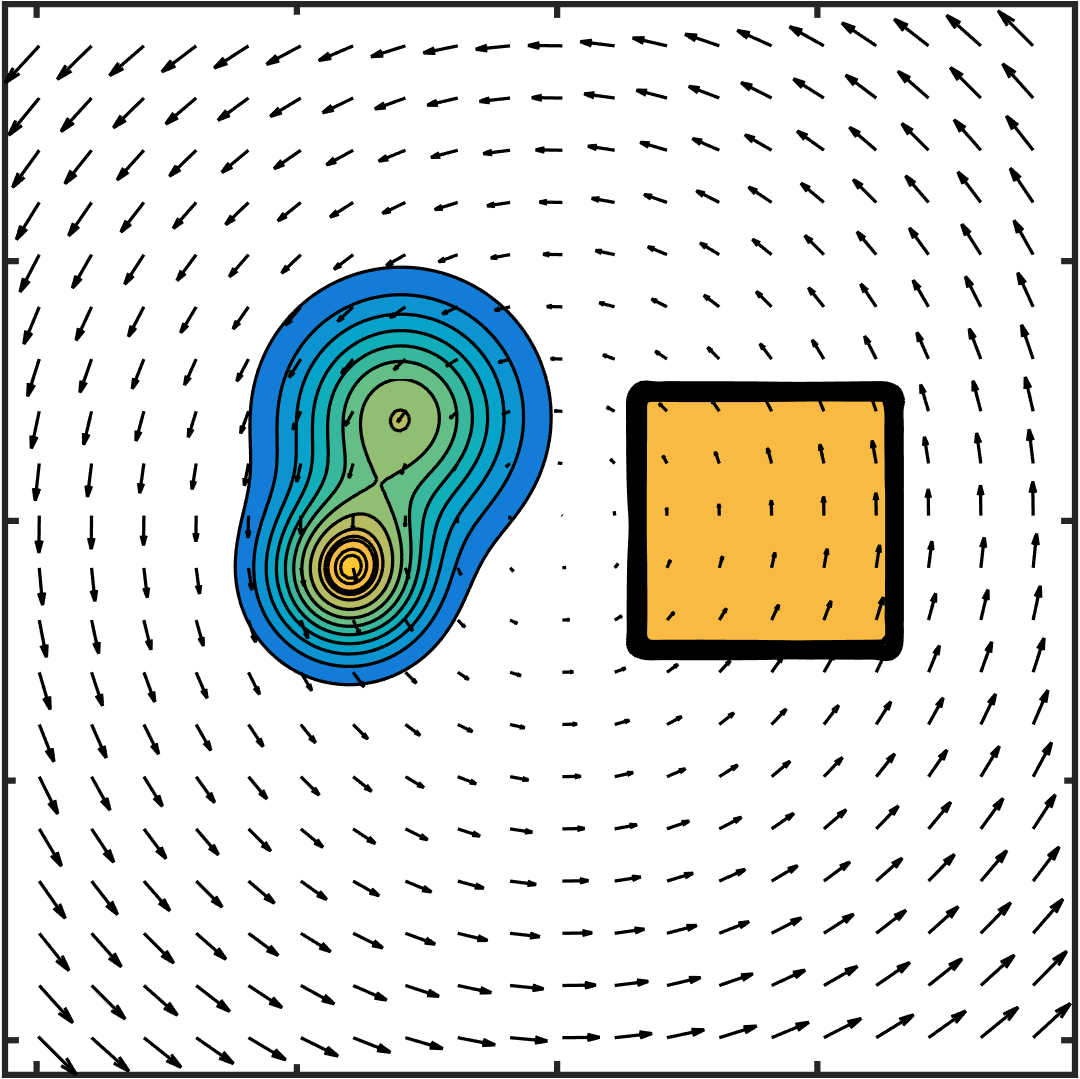}
\end{center}

\caption{Snapshots of the two-dimensional solid rotation test case, showing contours of the tracer $Q$ over a complete cycle (counter-clockwise from top-right). Contours are drawn at intervals of $0.1$ between ${-0.25}$ and ${+1.25}$. Results were obtained using the W-PQM/P$_4$E scheme on a $200\times 200$ grid.}

\end{figure}

\begin{figure}[t]

\label{figure_solid_rotation_results}

\begin{center}
\includegraphics[width=.275\textwidth]{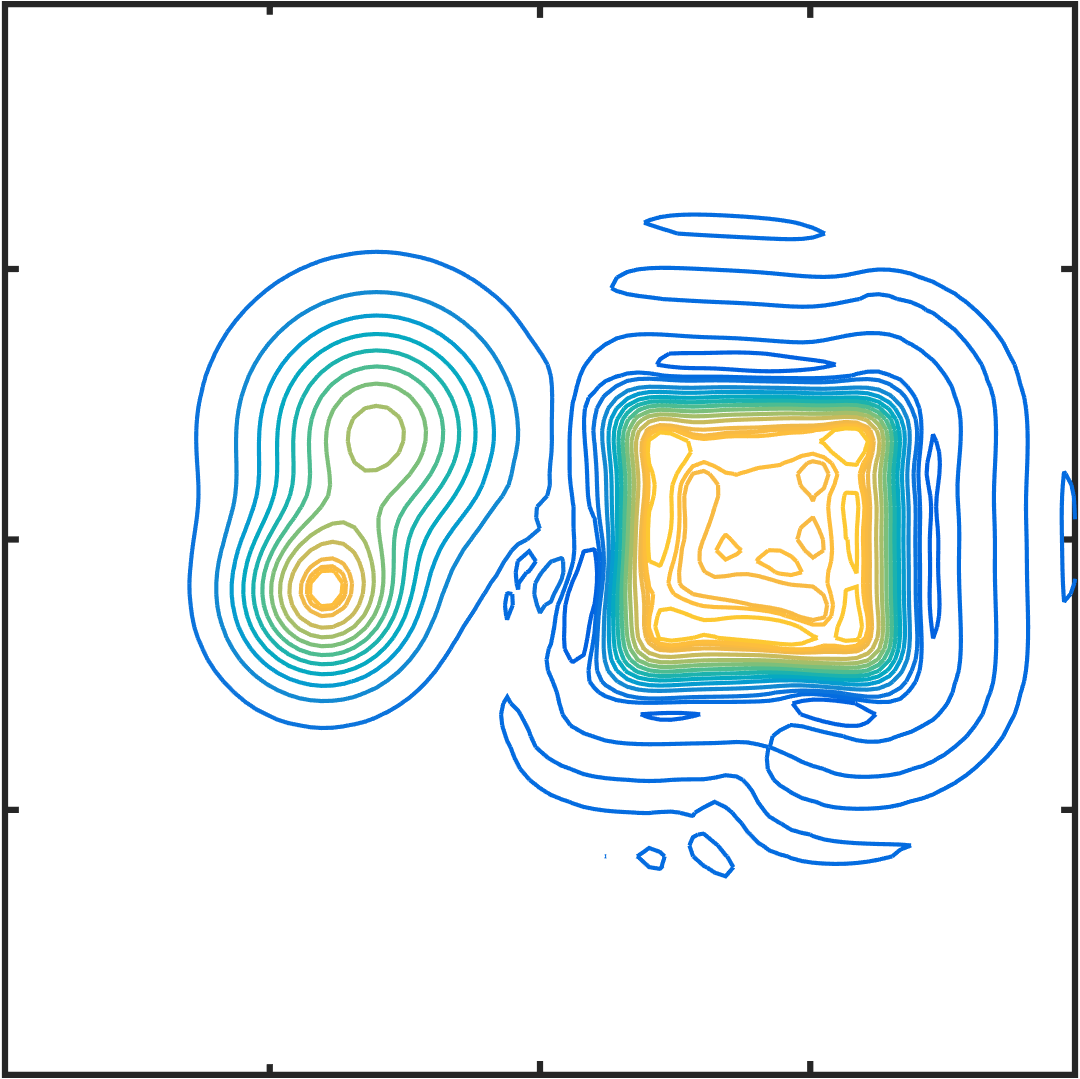}\quad
\includegraphics[width=.275\textwidth]{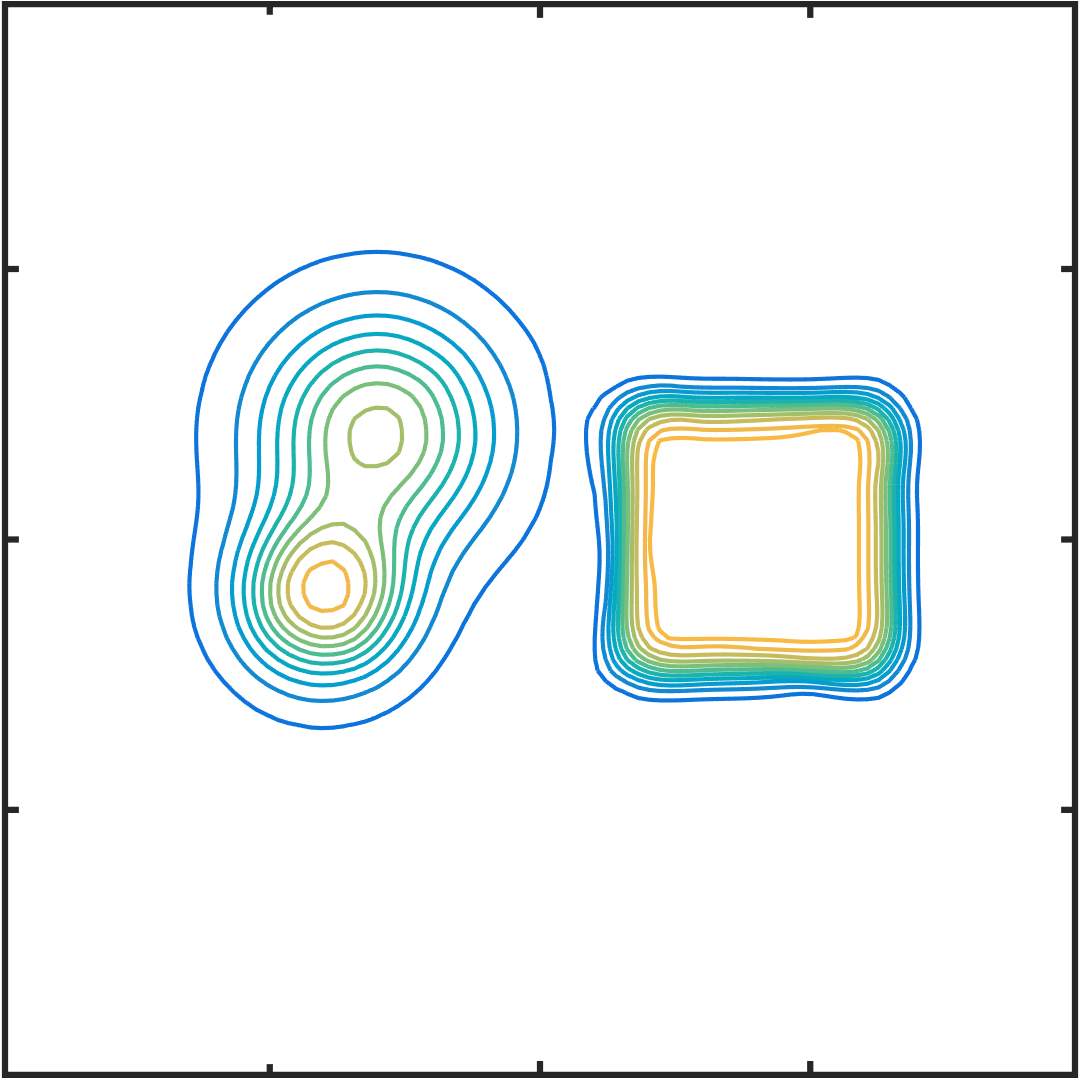}\quad
\includegraphics[width=.275\textwidth]{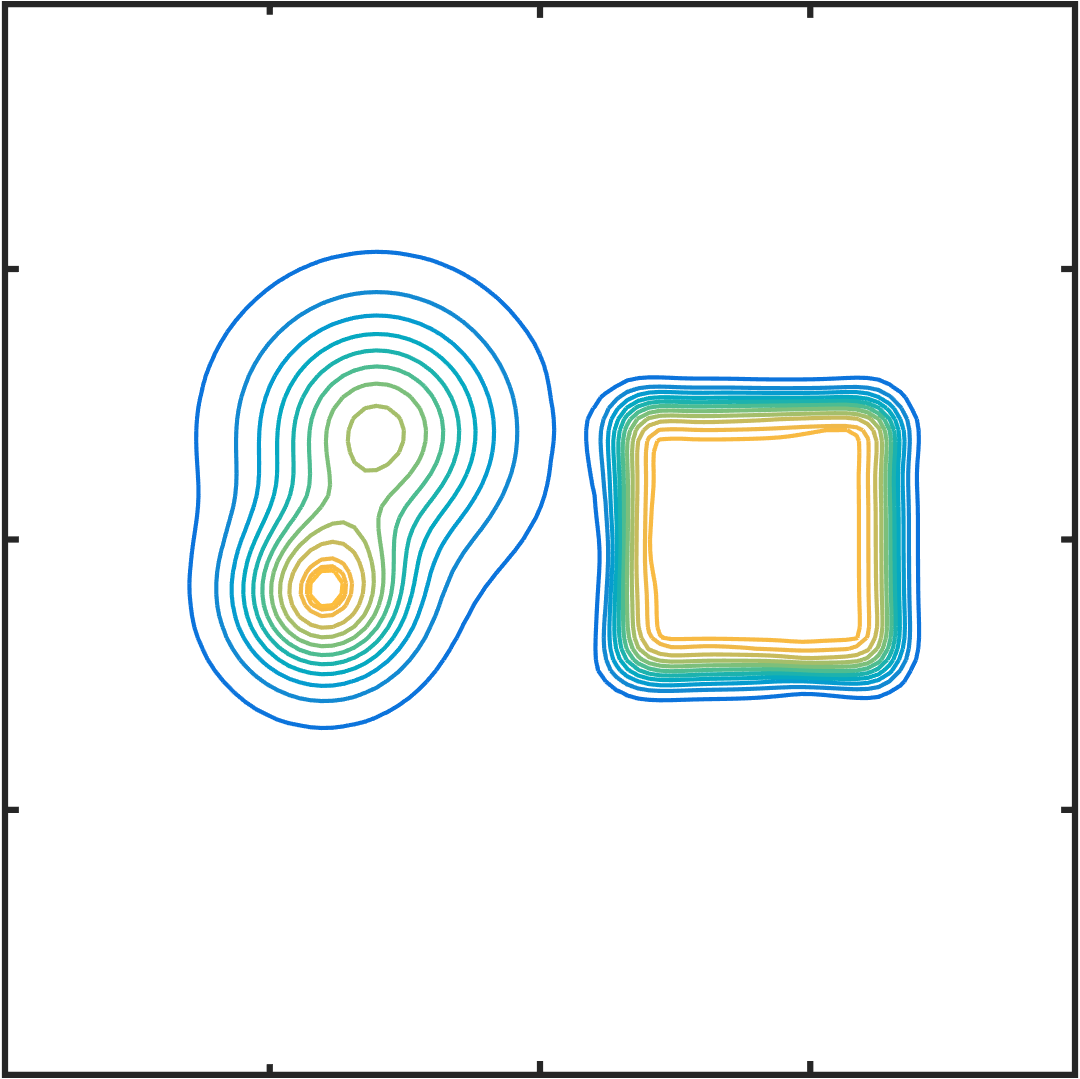}\\[2.25ex]
\includegraphics[width=.275\textwidth]{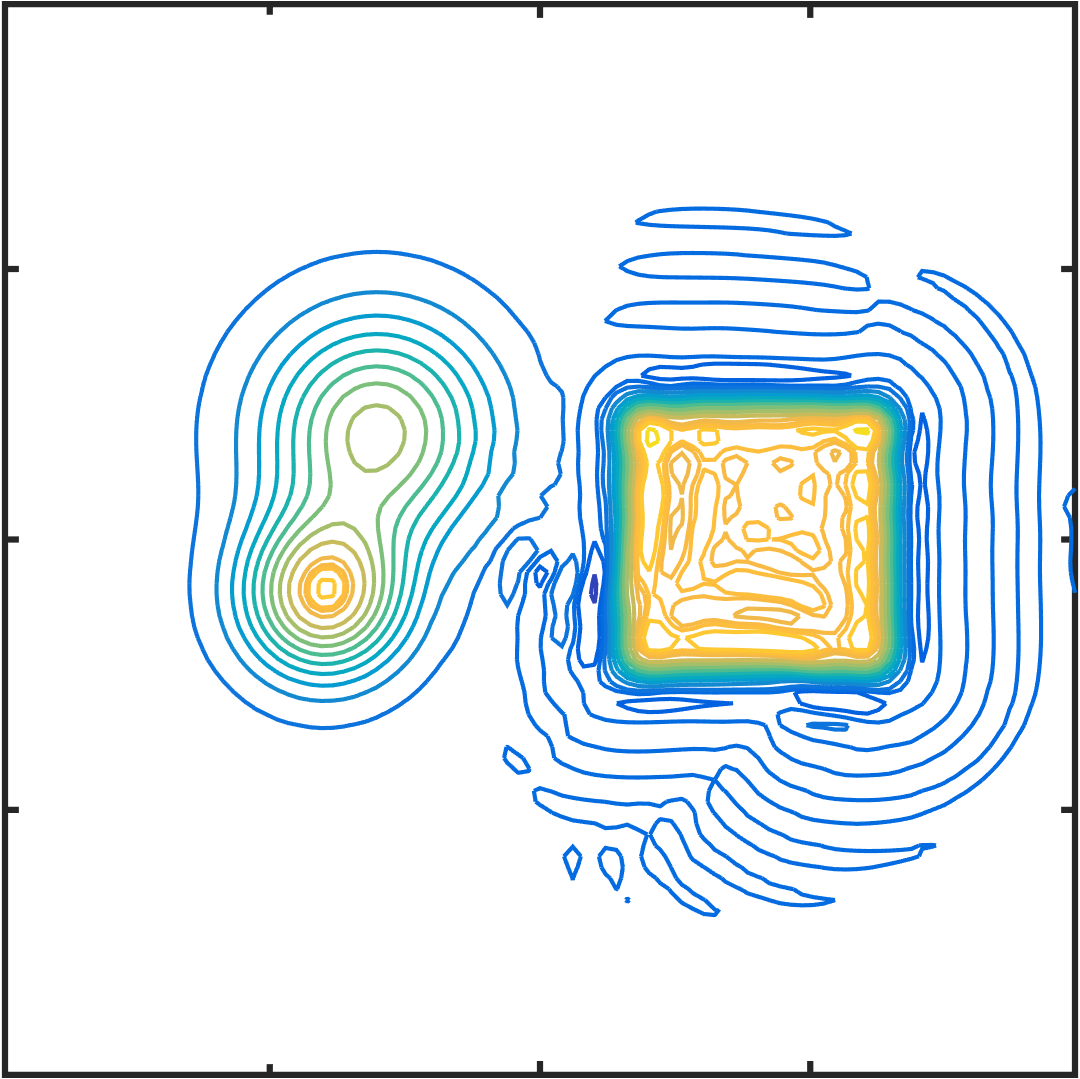}\quad
\includegraphics[width=.275\textwidth]{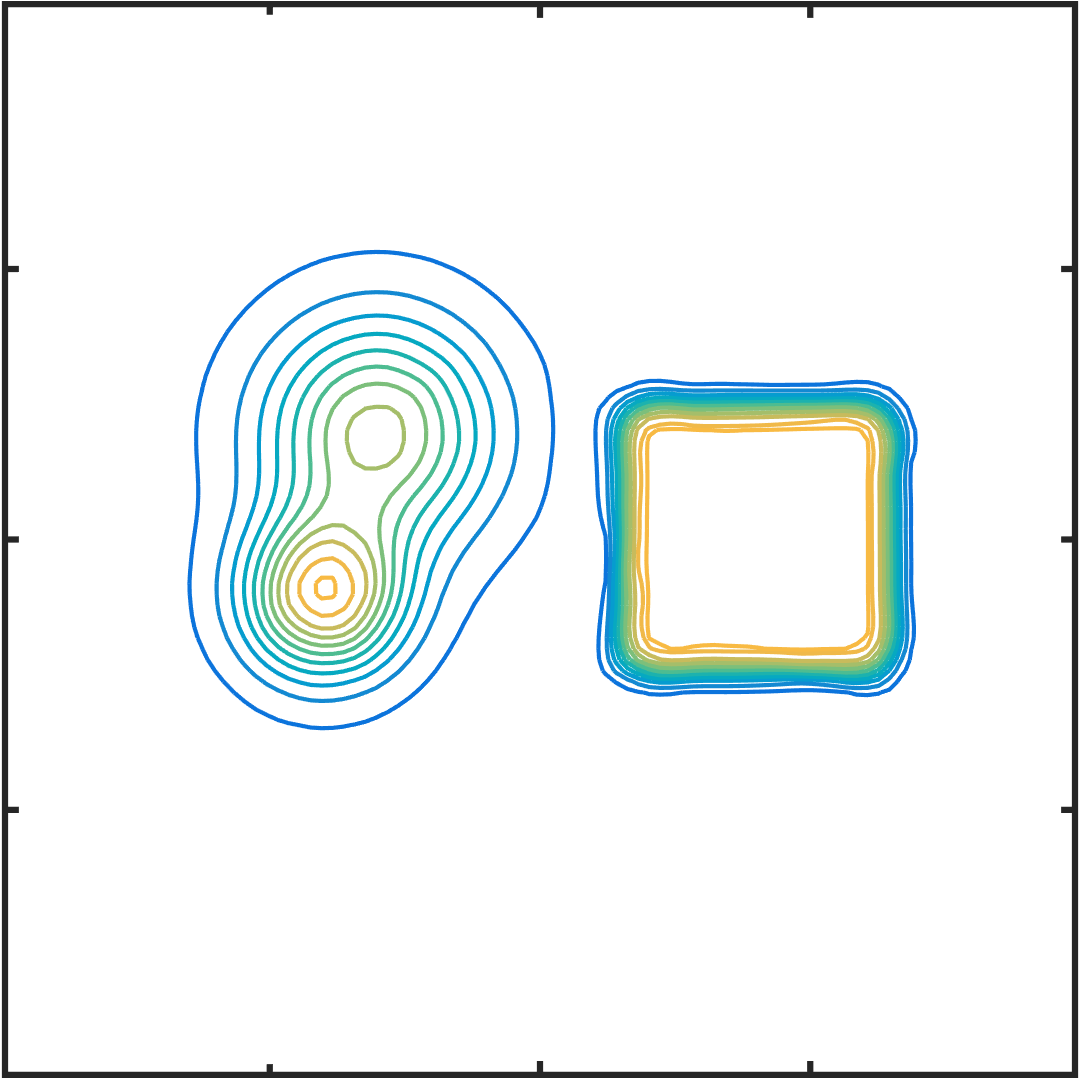}\quad
\includegraphics[width=.275\textwidth]{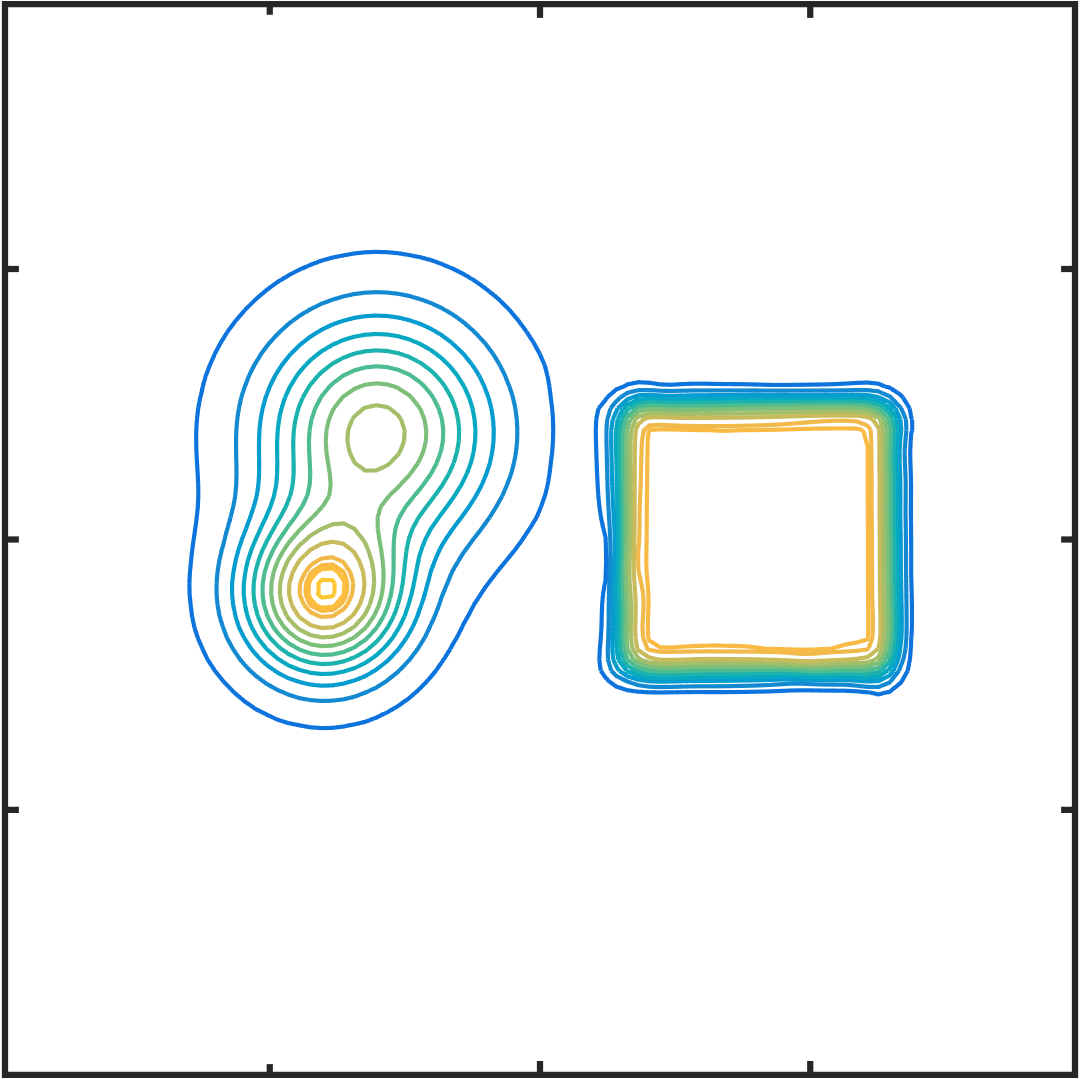}
\end{center}

\caption{Results for the solid rotation test case. Results are shown after four complete cycles $(t=8\pi)$. Contours are drawn at intervals of $0.1$ between ${-0.25}$ and ${+1.25}$ with additional levels also drawn at ${\pm 1\times 10^{-4}}$, ${+0.99}$ and ${+1.01}$. Results for all schemes were obtained using a $100\times 100$ grid.}

\end{figure}

The performance of the various reconstruction schemes was also assessed for two-dimensional scalar advection, with the direction-split semi-Lagrangian algorithm described in Section~\ref{section_lagrangian} used to integrate a series of benchmark problems. The following initial condition
\begin{eqnarray}
\label{eqn_profile_2d}
Q^{0} = \left\{
\begin{array}{cc}
{+1}, &
\text{if } \left(\frac{15}{100} \leq x \leq \frac{65}{100}\text{ and }{-\frac{25}{100}} \leq y \leq \frac{25}{100}\right) 
\\[4ex]
\begin{array}{r}
\operatorname{e}^{\left(-50\left((x+\frac{40}{100})^{2}-(y+\frac{10}{100})^{2}\right)\right)} \\[1ex]
+\frac{4}{5}\operatorname{e}^{\left(-25\left((x+\frac{30}{100})^{2}-(y-\frac{20}{100})^{2}\right)\right)}
\end{array}, & \text{ otherwise}
\end{array}
\right.
\end{eqnarray}
consisting of a pair of Gaussian profiles and a discontinuous `hat' function was utilised, providing a assessment of the various schemes for both smooth and discontinuous solution features. In the first test-case, the initial profile (\ref{eqn_profile_2d}) was subject to a solid-body rotational flow, given by
\begin{eqnarray}
\mathbf{u}(x,y) = - y\,\hat{\mathbf{x}} + x\,\hat{\mathbf{y}}.
\end{eqnarray}
The direction-split semi-Lagrangian presented in Section~\ref{section_lagrangian} was used to integrate the scalar advection equation (\ref{eqn_model_transport}) over four complete revolutions. This process is illustrated in Figure~\ref{figure_solid_rotation}. Importantly, note that solid-body rotation does not result in a deformation of $Q(x,y)$ over time. Consistent with the one-dimensional experiments presented previously, the solid-body rotation test was computed using the various combinations of the PPM and PQM reconstruction techniques, including both monotone and WENO-type slope-limiters. Additionally, a fully-unlimited reconstruction was also calculated. Contours of $Q(x,y)$ are presented in Figure~\ref{figure_solid_rotation_results}, with the unlimited, monotone, and WENO-based PPM schemes shown in the left column, and the associated PQM-based solutions shown on the right. All results were computed using a uniform $100 \times 100$ grid and a CFL number of 1. Results are calculated using the P$_2$E and P$_4$E edge-estimates for the PPM- and PQM-based schemes, respectively, though little variation was observed between the cell- and edge-centred formulations.

An inspection of the contours shown in Figure~\ref{figure_solid_rotation_results} demonstrates that both the monotone and WENO-type slope-limiting strategies for the PPM- and PQM-based semi-Lagrangian schemes result in acceptable solutions to the solid-body rotation problem. As expected, schemes based on the unlimited reconstructions are seen to incorporate strong non-physical oscillations in the neighbourhood of the discontinuous hat profile. Both the monotone and WENO-based schemes, on the other hand, result in essentially oscillation-free solutions. Further inspection of the contour plots shown in Figure~\ref{figure_solid_rotation_results} confirms that many of the conclusions drawn previously for the one-dimensional experiments are applicable -- that the W-PPM and W-PQM reconstructions preserve high-order accuracy near smooth local extrema while also suppressing spurious oscillations near sharp features. Additionally, it can be seen that the M-PPM and M-PQM schemes result in a significant flattening of the smooth peaks in the profile. Overall, it is clear that the W-PQM based scheme leads to the most accurate solution, recovering a near-exact representation of the smooth features in the profile.

\subsection{Shearing flow on the sphere}

\medskip

\begin{figure}[t]

\label{figure_shearing}

\begin{center}
\includegraphics[width=.455\textwidth]{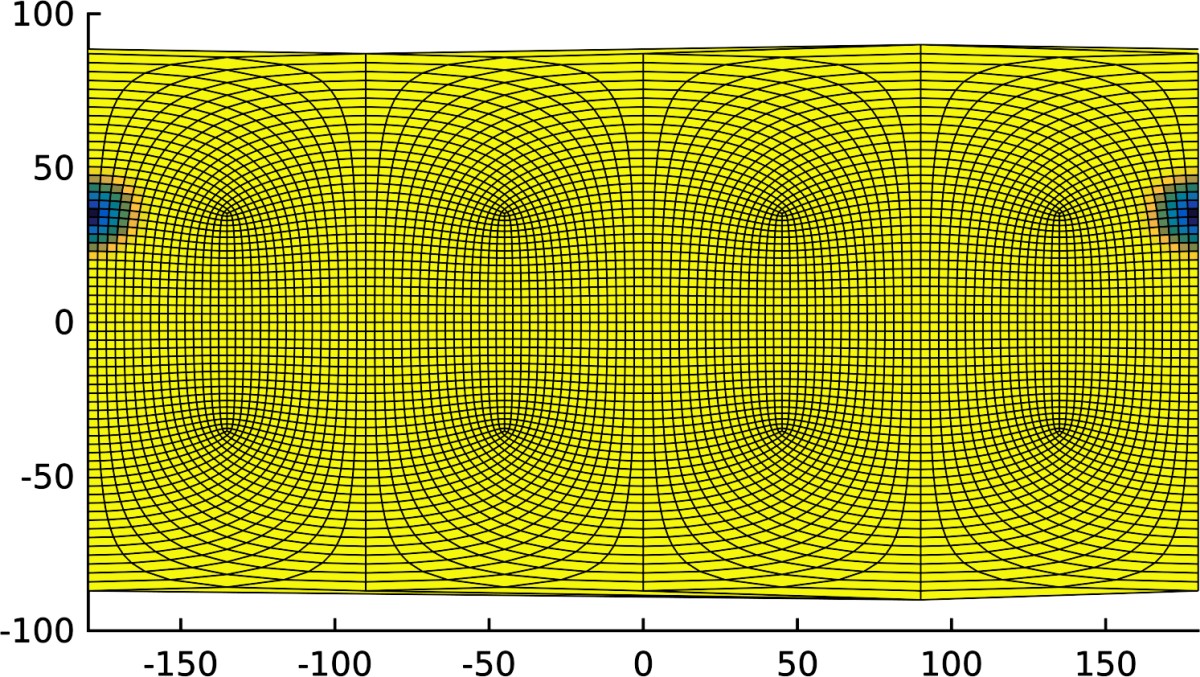}\quad
\includegraphics[width=.455\textwidth]{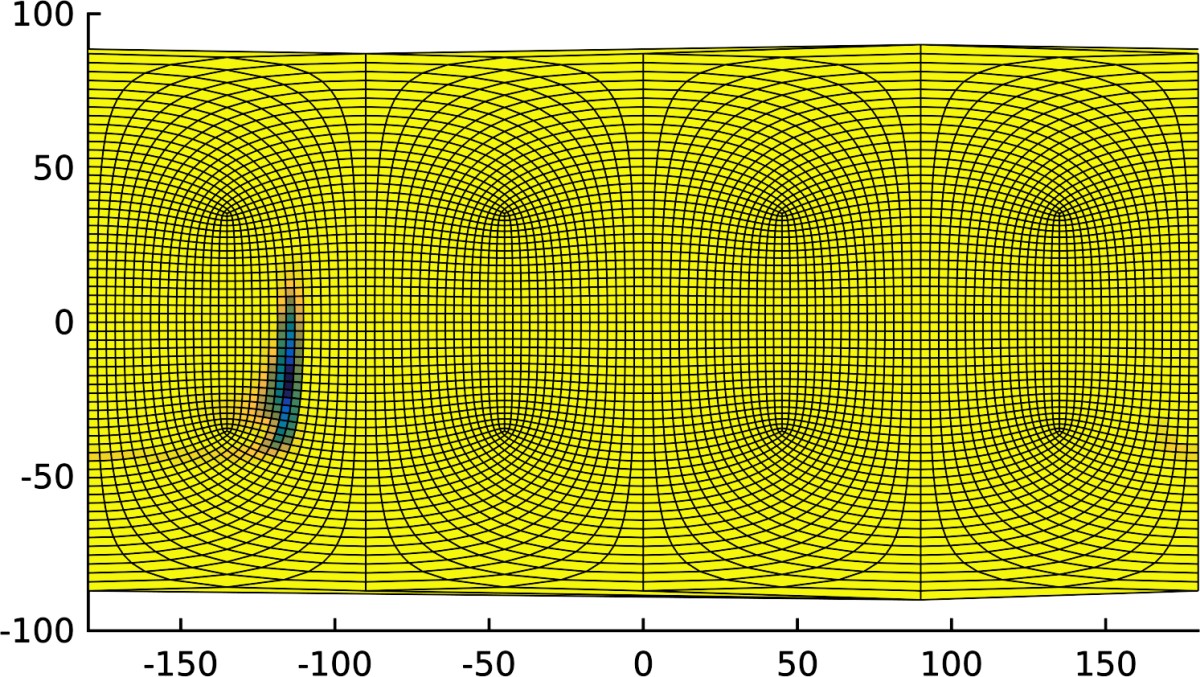}\\[2.25ex]
\includegraphics[width=.455\textwidth]{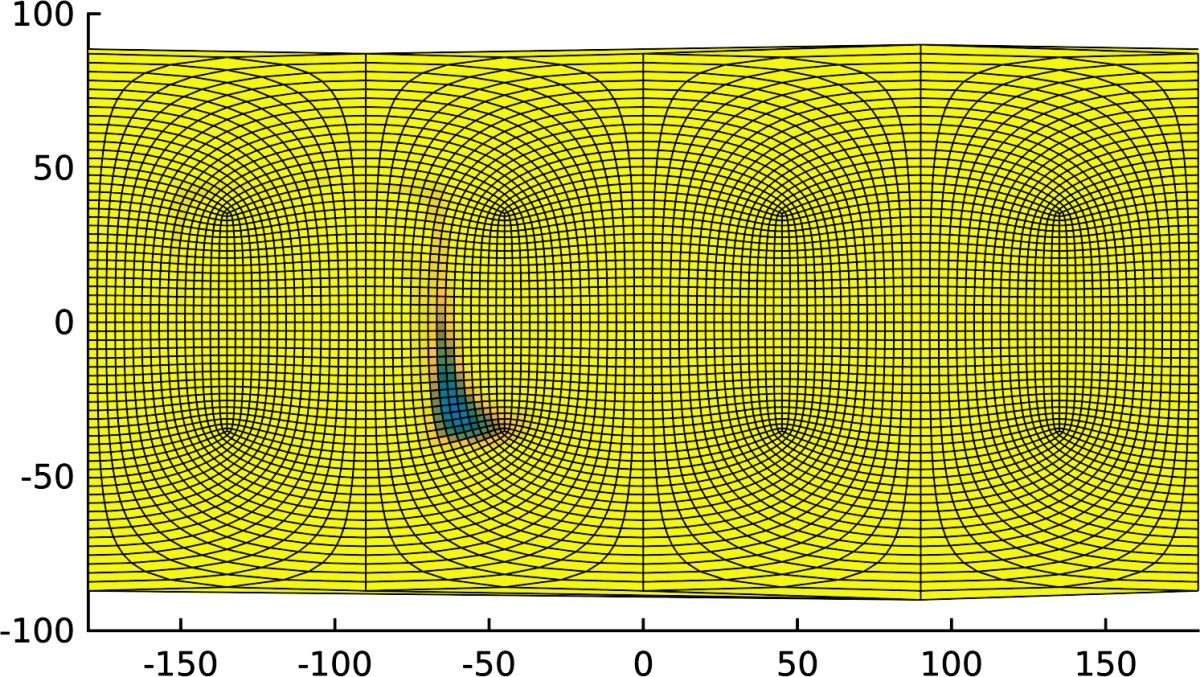}\quad
\includegraphics[width=.455\textwidth]{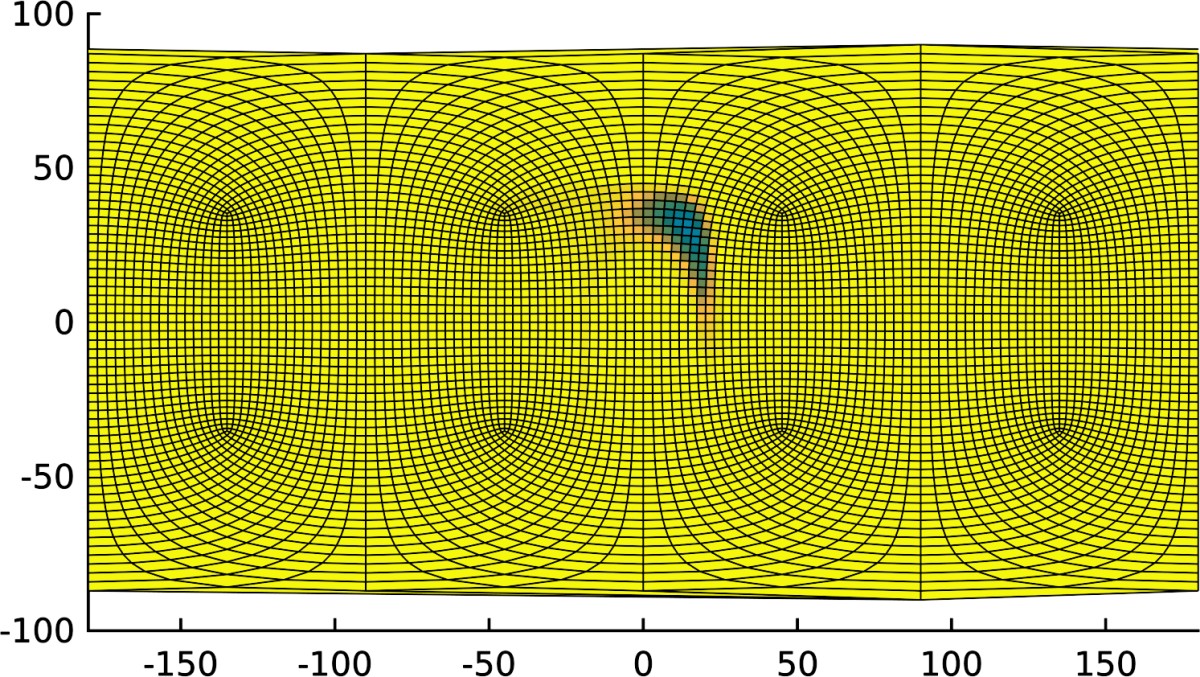}\\[2.25ex]
\includegraphics[width=.455\textwidth]{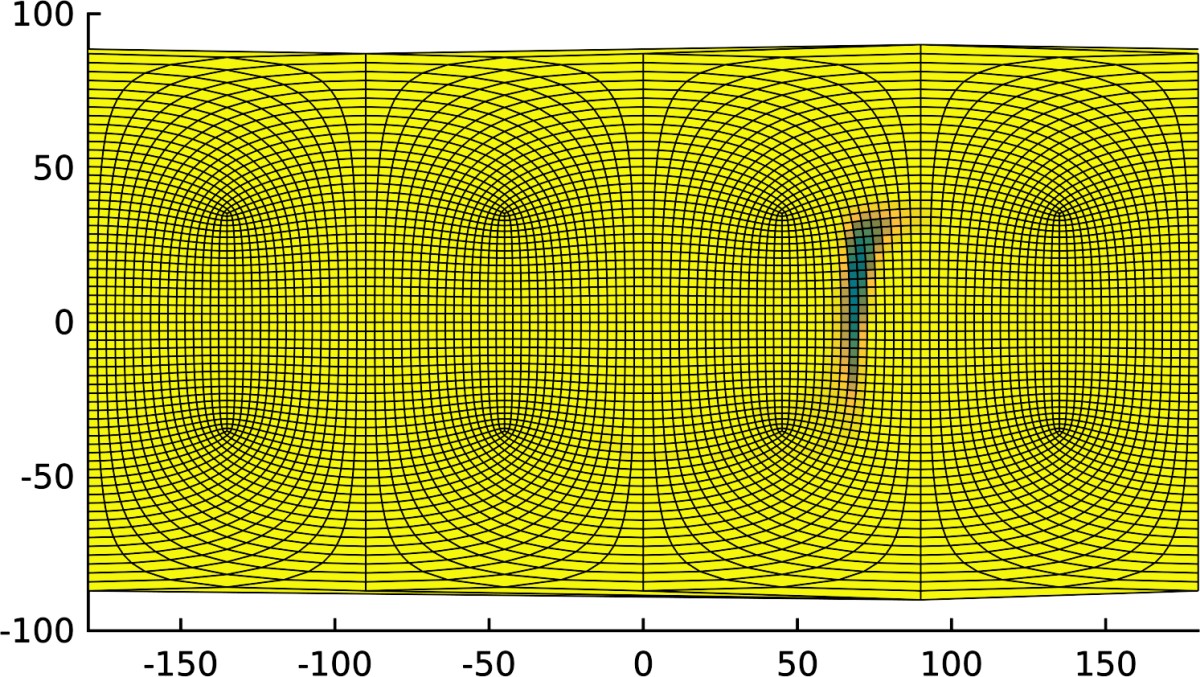}\quad
\includegraphics[width=.455\textwidth]{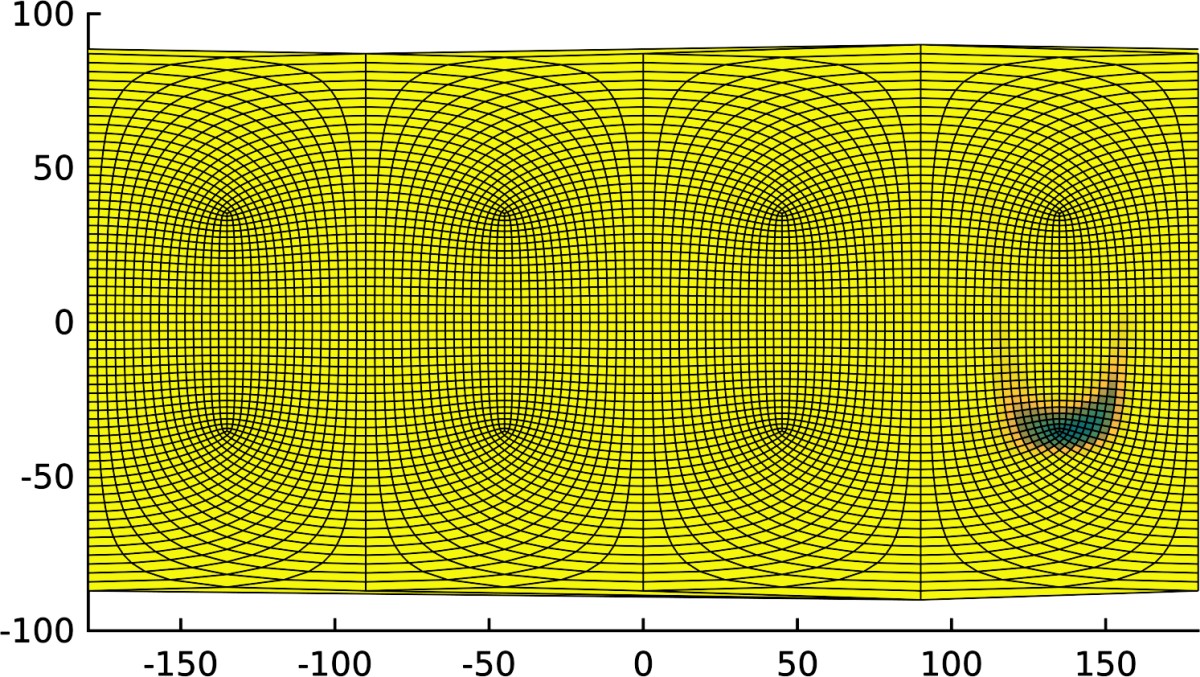}
\end{center}

\caption{Snapshots of the shearing test case on the cubed-sphere grid, showing the evolution of a tracer $Q$ over time. Results were obtained using the W-PQM/P$_5$E scheme on a conformal cubed-sphere grid, as implemented in the MITgcm. Each cube face consists of a $32 \times 32$ curvilinear grid. }

\end{figure}

The performance of the high-order advection schemes was also assessed using a more challenging problem, in which a non-uniform circulatng flow on the surface of a sphere was used to advect a passive tracer. In this test-case, the high-order PPM- and PQM-based reconstruction methods were implemented within the MITgcm \citep{marshall1997finite} -- a finite-volume type general-circulation model used for planetary climate studies. Consistent with the approach described in Section~\ref{section_lagrangian}, multi-dimensional advection in the MITgcm is achieved using a direction-splitting approach on a logically-rectangular cubed-sphere grid. The velocity field in this test-case is given by  
\begin{eqnarray}
\label{eqn_velocity_shearing}
\mathbf{u}(\psi,\theta) = \left(A \cos(\alpha\psi) \sin(\beta\theta) + \gamma\right)\,\hat{\mathbf{\psi}} - B \sin(\alpha\psi) \cos(\beta\theta)\,\hat{\mathbf{\theta}},
\end{eqnarray}
with $\alpha=4$, $\beta=2$, $\gamma=2$, $A = 8$ and $B = 16$. The initial tracer profile consists of a Gaussian hat, centred at $35^\circ$N and $180^\circ$E. In this problem, the tracer distribution is advected from West-to-East, following an oscillatory `snake'-like trajectory over the equator. In addition to the constant Easterly drift, the velocity field induces significant shear and deformation over time. See Figure~\ref{figure_shearing} for the time evolution of the tracer field. Due to the deformation of the profile and application to the curvilinear cubed-sphere grid, this benchmark represents a significantly more stringent test of the advection algorithms than the two-dimensional solid rotational flow presented previously. 

\begin{figure}[t]

\label{figure_shearing_results}

\begin{center}
\parbox{.455\textwidth}{\centering (i)} \quad \parbox{.455\textwidth}{\centering (ii)} \\
\includegraphics[width=.455\textwidth]{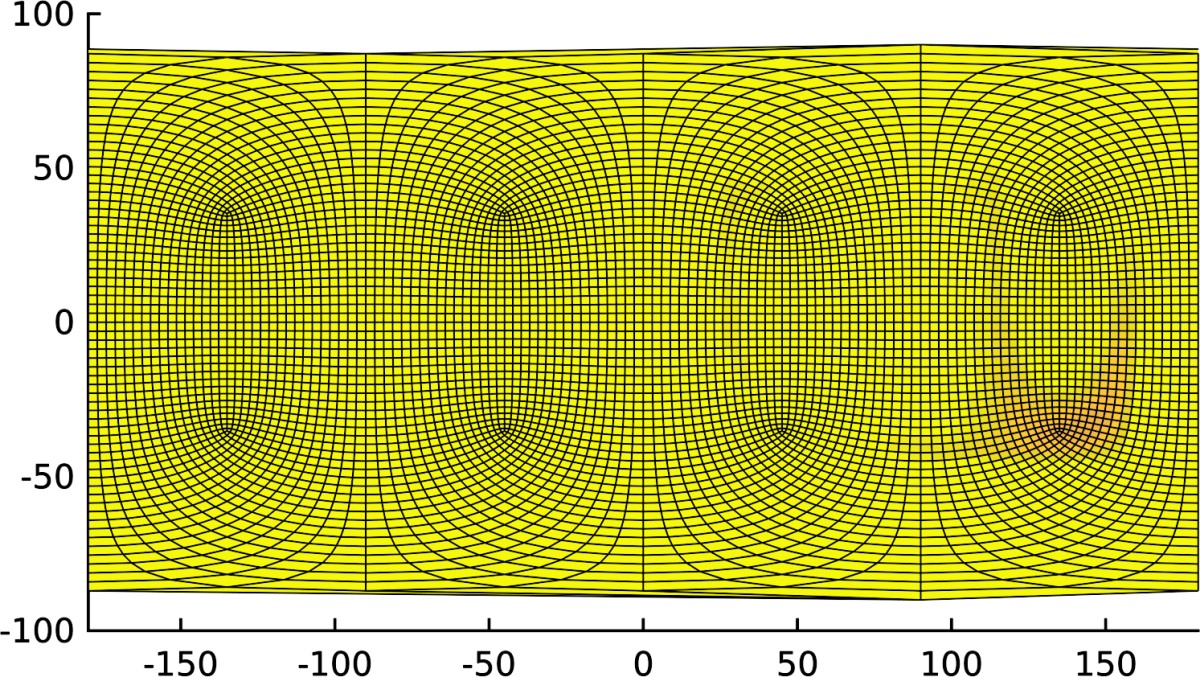}\quad
\includegraphics[width=.455\textwidth]{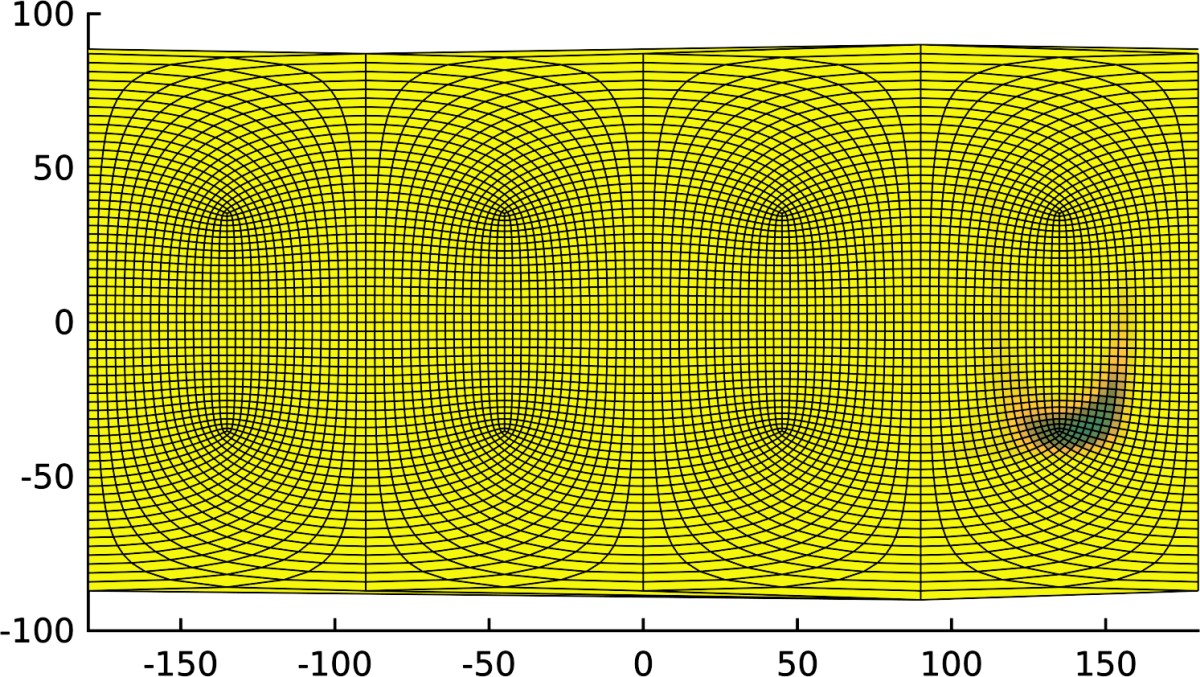}\\[2.25ex]
\parbox{.455\textwidth}{\centering (iii)} \quad \parbox{.455\textwidth}{\centering (iv)} \\
\includegraphics[width=.455\textwidth]{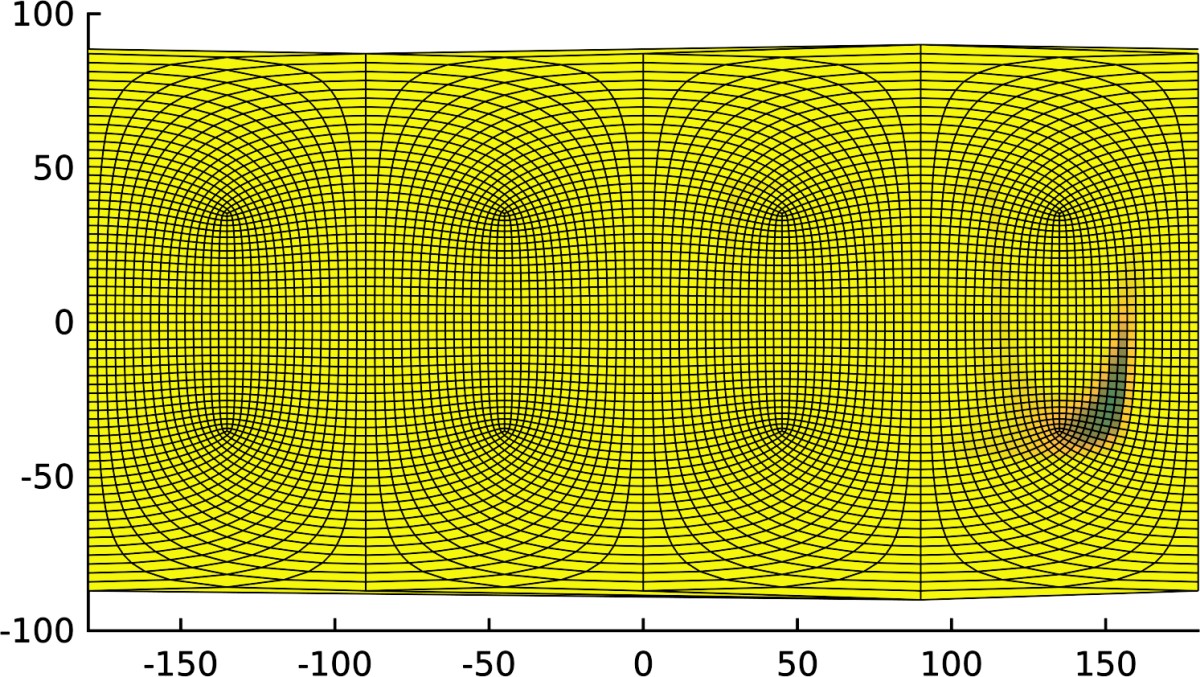}\quad
\includegraphics[width=.455\textwidth]{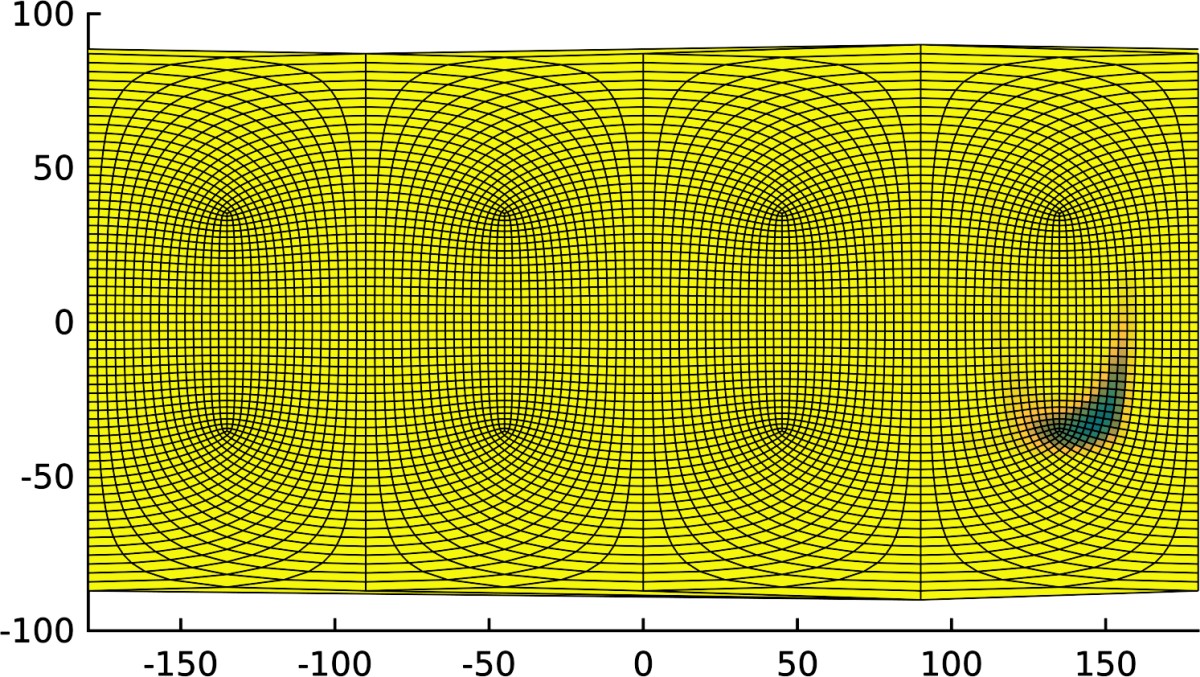}
\end{center}

\caption{Results for the shearing test case on the cubed-sphere grid, showing the final tracer distributions obtained using the advection schemes: (i) 3-DST, (ii) W-PPM/P$_3$E, (iii) OS7-MP, and (iv) W-PQM/P$_5$E.}

\end{figure}

\begin{figure}[t]

\label{figure_maxQ_2d}

\begin{center}
\includegraphics[width=.65\textwidth]{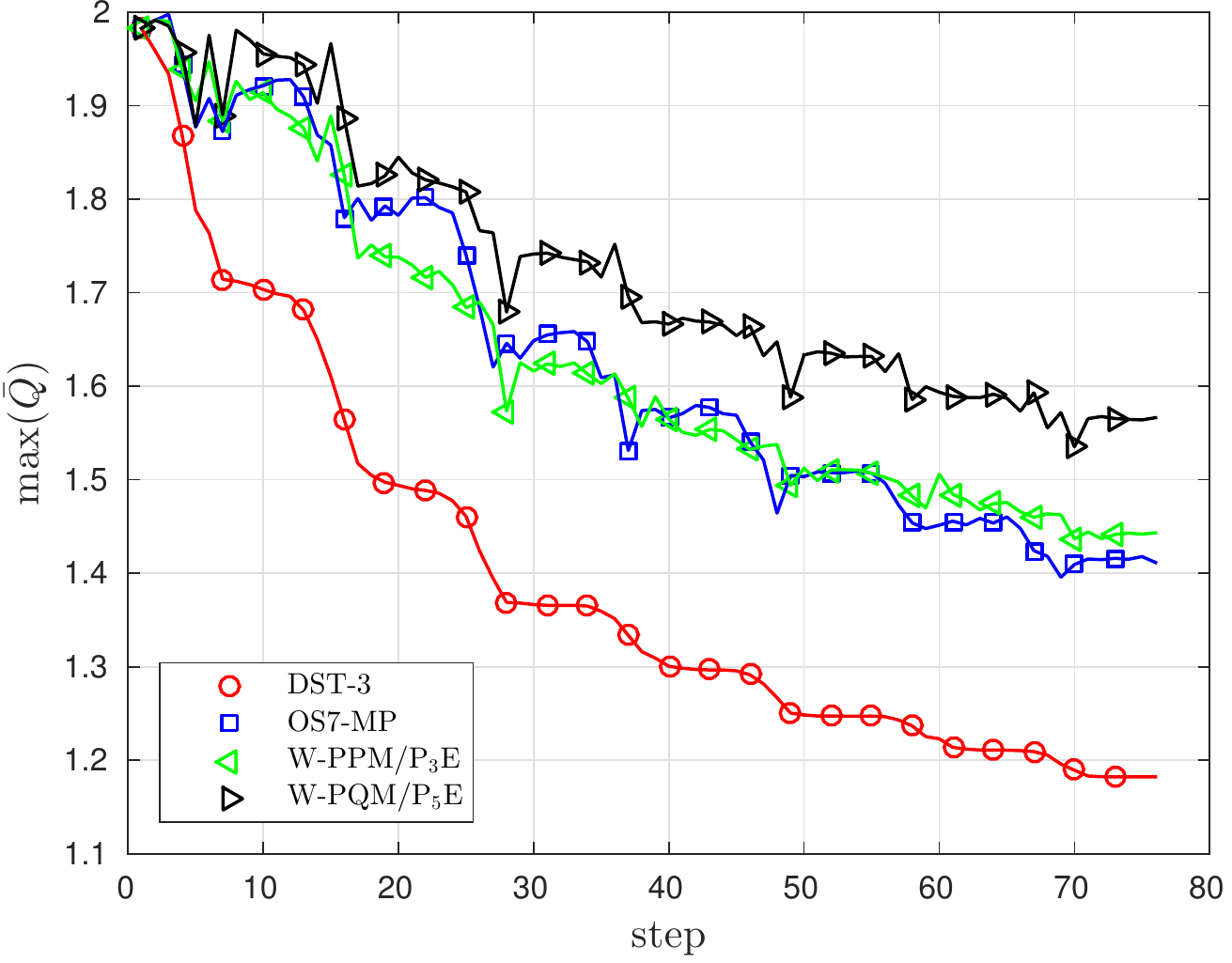}
\end{center}

\caption{Reduction in maximum tracer value over time in the shearing advection tests on the cubed-sphere grid, showing the variation amongst numerical advection schemes. Less reduction is better. The schemes include: (i) DST-3: a third-order accurate direct space-time method, (ii) OS7-MP: a 7th-order one-step method with a montonicity-preserving slope-limiter, (iii) W-PPM/W$_3$E: the WENO-limited piecewise parabloic method, and (iv) W-PQM/W$_5$E: the WENO-limited piecewise quartic method.}

\end{figure}

Results were computed using the WENO-based PPM and PQM interpolation schemes, in addition to a set of existing finite-volume type algorithms already implemented in the MITgcm. Specifically, the DST-3 \citep{HUNDSDORFER199535} (third-order accurate direct space-time with flux-limiting) and OS7-MP \citep{Daru2004563} (seventh-order accurate one-step method with monotonicty presvering slope-limiting) were included in the comparison, with the DST-3 scheme expected to offer similar performance to the PPM-type methods, and the OS7-MP scheme expected to compete with the PQM-type approaches. Results were computed using the P$_3$E and P$_5$E edge-estimates for the PPM- and PQM-based schemes, respectively. Contours for the final time-slice are shown in Figure~\ref{figure_shearing_results}, illustrating that, in a qualitative sense, there is good agreement between the schemes. All methods are found to result in smooth advection of the tracer, without appreciable under- or over-shoots. Based on the magnitude of the contour plot, it is clear that the DST-3 scheme is significantly more diffusive than the other methods. The magnitude of spurious numerical diffusion for each scheme was analysed by tracking the maximum tracer value $\max(\bar{Q}_{i,j})$ at each step of the simulation, with less diffusive methods showing a better preservation of the initial maximum. These trends are shown in Figure~\ref{figure_maxQ_2d}, demonstrating that (i) the DST-3 scheme is indeed the most diffusive of the methods studied, (ii) the W-PPM/P$_3$E and OS7-MP schemes lead to similar behaviour for this test-case, and (iii) the W-PQM/P$_5$E scheme is the least diffusive scheme included in the current study. Overall, it is clear that the W-PQM/P$_5$E scheme resulted in the most accurate solution.

\section{Conclusions}

A new slope-limiting procedure for the piecewise parabolic and piecewise quartic reconstruction methods (PPM and PQM) has been developed, based on a variation of the well-known WENO methodology. In contrast to conventional monotonicity preserving formulations, the new WENO-type slope-limiter has been designed to reconstruct smooth, essentially non-oscillatory polynomial profiles based on a non-linear combination of the \textit{natural} and \textit{monotone-limited} PPM and PQM polynomials. Such a formulation is designed to preserve high-order accuracy at smooth and well-resolved local extrema. Based on a comprehensive set of one-dimensional benchmarks, it has been demonstrated that the new WENO-limited PPM and PQM interpolants are effective in practice -- able to preserve high-order accuracy at smooth local extrema whilst simultaneously suppressing spurious oscillations in the neighbourhood of sharp and/or poorly resolved features. Specifically, it has been shown that the new W-PPM and W-PQM interpolants exhibit smooth third- and fifth-order spatial accuracy when the underlying data is sufficiently smooth. Additionally, it has been demonstrated that these methods are efficient in practice, requiring only a 10--20\% increase in computational effort for substantial reductions in numerical error. Based on comparisons of relative efficiency, it has been shown that the new W-PQM interpolant offers superior overall performance, while the W-PPM reconstruction scheme outperforms the associated monotone techniques by a smaller margin.

The new PPM and PQM reconstruction schemes were subsequently used to construct a high-order accurate Arbitrary-Lagrangian-Eulerian algorithm for advective transport problems. This method was extended to handle multi-dimensional problems through a direction-splitting approach using the MITgcm. These methods were applied to a set of two-dimensional scalar transport problems, including a solid rotational flow, and a shearing flow on the surface of the sphere. Consistent with the one-dimensional analysis, it was shown that the ALE formulations based on the new W-PPM and W-PQM interpolants were highly effective in practice, providing high-order accurate and essentially oscillation-free scalar advection. Performance for the new methods was shown to be superior to that of other well-known finite-volume type approaches. In contrast to conventional monotonicity preserving methods, it has been demonstrated that significant improvements in accuracy can be achieved when using the new WENO-based schemes when the underlying solution features are sufficiently smooth.

\section*{Acknowledgements}

This work was supported by a NASA-GISS / MIT cooperative research agreement.

\appendix

\section{Explicit edge-estimates}
\label{edge_appendix}

Given a uniform mesh-spacing $h$, explicit formulations for the P$_2$E, P$_3$E, P$_4$E and P$_5$E edge-estimates can be obtained. Specifically, estimates are given by
\begin{itemize}
\item P$_3$E scheme:
\begin{eqnarray}
q_{i+\frac{1}{2}} = 
-\frac{ 1}{12}q_{i-2}
+\frac{ 7}{12}q_{i-1}
+\frac{ 7}{12}q_{i}
-\frac{ 1}{12}q_{i+1}
\\[1ex]
\left(\frac{\partial q}{\partial x}\right)_{i+\frac{1}{2}} = 
 \frac{1}{h}\left(
+\frac{ 1}{12}q_{i-2}
-\frac{15}{12}q_{i-1}
+\frac{15}{12}q_{i}
-\frac{ 1}{12}q_{i+1}
\right)
\end{eqnarray}
\item P$_5$E scheme:
\begin{eqnarray}
q_{i+\frac{1}{2}} = 
 \frac{ 1}{60}q_{i-3}
-\frac{ 8}{60}q_{i-2}
+\frac{37}{60}q_{i-1}
+\frac{37}{60}q_{i}
-\frac{ 8}{60}q_{i+1}
+\frac{ 1}{60}q_{i+2}
\\[1ex]
\left(\frac{\partial q}{\partial x}\right)_{i+\frac{1}{2}} = 
 \frac{1}{h}\left(
-\frac{ 1}{90}q_{i-3}
+\frac{ 5}{36}q_{i-2}
-\frac{49}{36}q_{i-1}
+\frac{49}{36}q_{i}
-\frac{ 5}{36}q_{i+1}
+\frac{ 1}{90}q_{i+2}
\right)
\end{eqnarray}
\item P$_2$E scheme:
\begin{eqnarray}
q_{i-\frac{1}{2}}^{+} = 
+\frac{ 2}{6}q_{i-1}
+\frac{ 5}{6}q_{i}
-\frac{ 1}{6}q_{i+1}
\\[1ex]
q_{i+\frac{1}{2}}^{-} = 
-\frac{ 1}{6}q_{i-1}
+\frac{ 5}{6}q_{i}
+\frac{ 2}{6}q_{i+1}
\\[1ex]
\left(\frac{\partial q}{\partial x}\right)_{i-\frac{1}{2}}^{+} = 
 \frac{1}{h}\left(
-q_{i-1}
+q_{i}
\right)
\\[1ex]
\left(\frac{\partial q}{\partial x}\right)_{i+\frac{1}{2}}^{-} = 
 \frac{1}{h}\left(
-q_{i}
+q_{i+1}
\right)
\end{eqnarray}
\item P$_4$E scheme:
\begin{eqnarray}
q_{i+\frac{1}{2}}^{-} =
-\frac{ 3}{60}q_{i-2} 
+\frac{27}{60}q_{i-1}
+\frac{47}{60}q_{i}
-\frac{13}{60}q_{i+1}
+\frac{ 2}{60}q_{i+2}
\\[1ex]
q_{i+\frac{1}{2}}^{+} =
+\frac{ 2}{60}q_{i-2} 
-\frac{13}{60}q_{i-1}
+\frac{47}{60}q_{i}
+\frac{27}{60}q_{i+1}
-\frac{ 3}{60}q_{i+2}
\\[1ex]
\left(\frac{\partial q}{\partial x}\right)_{i+\frac{1}{2}}^{-} = 
 \frac{1}{h}\left(
+\frac{ 1}{12}q_{i-2}
-\frac{15}{12}q_{i-1}
+\frac{15}{12}q_{i}
-\frac{ 1}{12}q_{i+1}
\right)
\\[1ex]
\left(\frac{\partial q}{\partial x}\right)_{i+\frac{1}{2}}^{+} = 
 \frac{1}{h}\left(
+\frac{ 1}{12}q_{i-1}
-\frac{15}{12}q_{i}
+\frac{15}{12}q_{i+1}
-\frac{ 1}{12}q_{i+2}
\right)
\end{eqnarray}
\end{itemize}

\section{Monotone slope-limiting for PPM}
\label{ppm_poly_appendix}

The PPM and PQM slope-limiters rely on an evaluation of the local left-, right- and centre-biased piecewise linear slope estimates (\citet{white2008high})
\begin{eqnarray}
\sigma_{\text{R}} = 2\left(\frac{\bar{q}_{i+1}-\bar{q}_{i}}{h_{i}}\right),\quad
\sigma_{\text{C}} = 2\left(\frac{\bar{q}_{i+1}-\bar{q}_{i-1}}{h_{i-1}+2h_{i}+h_{i+1}}\right),\quad
\sigma_{\text{L}} = 2\left(\frac{\bar{q}_{i}-\bar{q}_{i-1}}{h_{i}}\right).
\end{eqnarray}
A limited linear slope $\sigma$ is defined for each grid-cell via the well-known minmod function, denoted here $\text{minmod}\left(\cdot\right)$, such that
\begin{eqnarray}
\label{eqn_pls}
\sigma = \text{minmod}\left(\sigma_{\text{C}},\text{minmod}\left(\sigma_{\text{R}}
,\sigma_{\text{L}}\right)\right)
\end{eqnarray}
where
\begin{eqnarray}
\text{minmod}(a,b) = \left\{
\begin{array}{ll}
a, & \text{if } \left(ab > 0\right) \text{ and } \left(|a| \leq |b|\right), \\[\matspace]
b, & \text{if } \left(ab > 0\right) \text{ and } \left(|b| \leq |a|\right), \\[\matspace]
0, & \text{otherwise }
\end{array}
\right.
\end{eqnarray}

\subsection{Limiting edge-estimates}

\medskip

Firstly, local extrema are detected and subsequently \textit{flattened}, such that $q_{\text{L}}=q_{\text{R}}=\bar{q}$. This process imposes piecewise constant profiles in the affected grid-cells. Local extrema correspond to local peaks or troughs in the cell-mean distribution, and are detected when
\begin{eqnarray}
\label{eqn_extrema}
(\bar{q}_{i+1}-\bar{q}_{i})(\bar{q}_{i}-\bar{q}_{i-1})\leq 0.
\end{eqnarray}
Following the flattening of local extrema, the boundedness of edge-values is enforced. An edge value is unbounded if it lies outside the range of adjacent cell-mean values. In such cases, the edge-value estimate is replaced with a slope-limited linear interpolation. The bounded edge-value estimates $\tilde{q}_{\text{L}}$ and $\tilde{q}_{\text{R}}$ can be expressed as
\begin{eqnarray}
\label{eqn_edge_bounds}
\tilde{q}_{\text{L}} = \left\{
\begin{array}{ll}
\bar{q}_{i}-\tfrac{1}{2}h_{i}\sigma, & \text{if } \left((\bar{q}_{i}-q_{\text{L}})(q_{\text{L}}-\bar{q}_{i-1}) \leq 0\right), \\[\matspace]
q_{\text{L}}, & \text{otherwise}
\end{array}
\right.
\\[2ex]
\tilde{q}_{\text{R}} = \left\{
\begin{array}{ll}
\bar{q}_{i}+\tfrac{1}{2}h_{i}\sigma, & \text{if } \left((\bar{q}_{i+1}-q_{\text{R}})(q_{\text{R}}-\bar{q}_{i}) \leq 0\right), \\[\matspace]
q_{\text{R}}, & \text{otherwise}
\end{array}
\right.
\end{eqnarray}
where use of the limited linear slope $\sigma$ ensures that $\tilde{q}_{\text{L}}$ and $\tilde{q}_{\text{R}}$ lie between the adjacent cell-mean values.

\subsection{Limiting grid-cell profiles}

\medskip

At this stage, local extrema have been flattened and edge-value estimates are bounded, but local monotonicity can still be violated due to the presence of local turning points in the grid-cell profiles $Q(\xi)$. Following \citet{colella1984piecewise}, the coefficients of the grid-cell profiles can be modified to move any internal turning points onto the closest grid-cell boundary. Recalling the form of the PPM interpolant (\ref{eqn_ppm}), the location of the turning point can be expressed as
\begin{eqnarray}
\label{eqn_ppm_turn}
\xi_{\text{T}} = -\frac{1}{2}\frac{\alpha_{1}}{\alpha_{2}}.
\end{eqnarray}
When $\xi_{\text{T}}$ is internal to a grid-cell, such that $\xi_{\text{T}}\in{[{-1},{+1}]}$, the opposing edge-value estimate is modified to move $\xi_{\text{T}}$ onto the nearest grid-cell edge
\begin{eqnarray}
\label{eqn_ppm_edge_value_soln_1}
\text{if }\left(\xi_{\text{T}}\in{[{-1},{+0}]}\right), \quad q_{\text{R}}\leftarrow 3\bar{q} - 2 q_{\text{L}}\\ 
\text{if }\left(\xi_{\text{T}}\in{[{+0},{+1}]}\right), \quad q_{\text{L}}\leftarrow 3\bar{q} - 2 q_{\text{R}}
\label{eqn_ppm_edge_value_soln_2}
\end{eqnarray}
Following these final modifications to the edge-values, the resulting slope-limited PPM interpolant is guaranteed to enforce exact monotonicity.

\section{Monotone slope-limiting for PQM}
\label{pqm_poly_appendix}

\subsection{Limiting edge-estimates}

\medskip

Consistent with the PPM slope-limiting formulation presented previously, the edge-value estimates are first modified to \textit{flatten} any local extrema, and to satisfy local boundedness constraints. This process is identical to that described previously via expressions (\ref{eqn_extrema}) and (\ref{eqn_edge_bounds}). Additionally, the consistency of the edge-slope estimates are checked against a local piecewise linear approximation. Specifically
\begin{eqnarray}
{\tilde{q}_{\text{R}}}' = \left\{
\begin{array}{ll}
\sigma, & \text{if }\left(\sigma {q_{\text{R}}}'\leq 0\right), \\[\matspace]
 {q_{\text{R}}}', & \text{otherwise}
\end{array}
\right. \quad
{\tilde{q}_{\text{L}}}' = \left\{
\begin{array}{ll}
\sigma, & \text{if }\left(\sigma {q_{\text{L}}}'\leq 0\right), \\[\matspace]
 {q_{\text{L}}}', & \text{otherwise}
\end{array}
\right.
\end{eqnarray}
where $\sigma$ is the slope-limited piecewise linear slope defined in (\ref{eqn_pls}).

\subsection{Limiting grid-cell profiles}

\medskip

At this stage, local extrema have been flattened, edge-value estimates bounded and edge-slope estimates modified for consistency, but local monotonicity can still be violated due to the presence of local turning points in the grid-cell profiles $Q(\xi)$. Following \citet{white2008high}, the coefficients of the grid-cell profiles are modified to move any internal \textit{inconsistent} inflexion points onto a grid-cell boundary. White and Adcroft have shown that such a constraint guarantees that local monotonicity is enforced. Recalling the form of the PQM interpolant (\ref{eqn_pqm}), the location of the inflexion points can be expressed as the solution to the following quadratic equation
\begin{eqnarray}
\label{eqn_pqm_quadratic}
12\alpha_{4}\xi_{\text{I}}^{2} + 6\alpha_{3}\xi_{\text{I}} + 2\alpha_{2} = 0.
\end{eqnarray}
When there exists an $\xi_{\text{I}}$ that is internal to the grid-cell and when the corresponding slope ${Q\left(\xi_{\text{I}}\right)}'$ is locally inconsistent, such that $\sigma {Q\left(\xi_{\text{I}}\right)}' \leq 0$, additional modifications to the edge-slope, and, possibly, edge-value estimates are required. These modifications are accomplished in two stages. 

\subsection{Modified edge-slope estimates}

\medskip

A modification of both the left and right edge-slope estimates is first attempted. This is done in order to preserve the $C_{0}$ continuity of the interpolant where possible. Following \citet{white2008high}, the inflexion points $\xi_{I}$ are moved onto the cell boundary associated with a smaller one-sided linear slope estimate. Specifically
\begin{eqnarray}
\text{if }\left(|\sigma_{\text{L}}| < |\sigma_{\text{R}}|\right) \quad \xi_{\text{I}}^{*}\leftarrow -1, \quad\text{else}\quad \xi_{\text{I}}^{*}\leftarrow +1
\end{eqnarray}  
where $\sigma_{\text{L}}$ and $\sigma_{\text{R}}$ are the one-sided linear slopes defined in (\ref{eqn_pls}) and $\xi_{\text{I}}^{*}$ is the desired location of the inflexion points. Given a target position for the inflexion points, the solution to the quadratic (\ref{eqn_pqm_quadratic}) is used to calculate a set of modified edge-slopes. Specifically, considering that solutions to (\ref{eqn_pqm_quadratic_appendix}) can be expressed as
\begin{eqnarray}
\xi_{\text{I}}^{*} = \pm 1 = \frac{ -\alpha_{3} \pm \left( \alpha_{3}^{2} - 16\alpha_{4}\alpha_{2} \right)^{_{\frac{1}{2}}}}{4\alpha_{4}}
\end{eqnarray}
a pair of constraints on the coefficients $\alpha_{i}$ can be obtained, such that
\begin{eqnarray}
\label{eqn_pqm_mod_quadratic}
\alpha_{3} - 16\alpha_{4}\alpha{2} = 0, \quad \pm 4\alpha_{4} = -\alpha_{3}
\end{eqnarray}
where the first expression in (\ref{eqn_pqm_mod_quadratic}) requires that the inflexion points constitute a double root, while the second expression in (\ref{eqn_pqm_mod_quadratic}) moves the inflexion point onto $\xi_{\text{I}}^{*}$. Following further algebraic manipulations of (\ref{eqn_pqm_mod_quadratic}), and using the explicit \textsc{pqm} reconstruction coefficients given in (\ref{eqn_pqm_soln}), the modified edge-slope estimates can be expressed as 
\begin{eqnarray}
\label{eqn_pqm_edge_slope_soln}
\begin{array}{l}
\text{if }(\xi_{\text{I}}^{*} = {-1}),\\[\matspace]
\begin{bmatrix}
{q_{\text{R}}}' \\[1ex]
{q_{\text{L}}}'
\end{bmatrix} = 
\begin{bmatrix}
-5 &  \phantom{-}3  &  \phantom{-}2 \\[\matspace]
\phantom{-}\frac{5}{3} & -\frac{1}{3} & -\frac{4}{3}
\end{bmatrix}
\begin{bmatrix}
\bar{q} \\
q_{\text{R}} \\
q_{\text{L}}
\end{bmatrix},
\end{array}
\qquad
\begin{array}{l}
\text{if }(\xi_{\text{I}}^{*} = {-1}),\\[\matspace]
\begin{bmatrix}
{q_{\text{R}}}' \\[1ex]
{q_{\text{L}}}'
\end{bmatrix} = 
\begin{bmatrix}
-\frac{5}{3} &  \phantom{-}\frac{4}{3} & \phantom{-}\frac{1}{3} \\[\matspace]
\phantom{-}5 &  -2 & -3   
\end{bmatrix}
\begin{bmatrix}
\bar{q} \\
q_{\text{R}} \\
q_{\text{L}}
\end{bmatrix}
\end{array}
\end{eqnarray}
While the inflexion points are now guaranteed to lie on cell-edges, as per \citet{white2008high}, the resulting PQM interpolants may now contain an inconsistent edge slope. This situation can be remedied through further modifications to both the edge-value and edge-slope estimates.

\subsection{Modified edge-value estimates}

\medskip

The consistency of the modified PQM edge-slopes are checked, and further modifications are enqueued if inconsistencies are detected. Specifically, any inconsistent edge-slopes are set to zero, while the opposite edge-value and edge-slope estimates are also modified, such that
\begin{eqnarray}
\label{eqn_pqm_edge_value_soln_1}
\begin{array}{l}
\text{if }\left(\xi_{\text{I}}^{*} = {-1}\right)\text{ and } \left(\sigma q_{\text{L}}' \leq 0\right), \quad q_{\text{L}}' = 0, \\[1ex]
\begin{bmatrix}
q_{\text{R}} \\[1ex]
q_{\text{R}}'
\end{bmatrix} = 
\begin{bmatrix}
\mspc 5 &  -4 \\[\matspace]
\mspc 10 & -10
\end{bmatrix}
\begin{bmatrix}
\bar{q} \\
q_{\text{L}}
\end{bmatrix},
\end{array}
\qquad
\begin{array}{l}
\text{if }\left(\xi_{\text{I}}^{*} = {-1}\right)\text{ and } \left(\sigma q_{\text{R}}' \leq 0\right), \quad q_{\text{R}}' = 0, \\[1ex]
\begin{bmatrix}
q_{\text{L}} \\[1ex]
q_{\text{L}}'
\end{bmatrix} = 
\begin{bmatrix}
\mspc\frac{5}{2} &  -\frac{3}{2} \mspc \\[\matspace]
-\frac{5}{3} & \mspc\frac{5}{3}  \mspc 
\end{bmatrix}
\begin{bmatrix}
\bar{q} \\
q_{\text{R}}
\end{bmatrix}
\end{array}
\end{eqnarray}
\begin{eqnarray}
\label{eqn_pqm_edge_value_soln_2}
\begin{array}{l}
\text{if }\left(\xi_{\text{I}}^{*} = {+1}\right)\text{ and } \left(\sigma q_{\text{L}}' \leq 0\right), \quad q_{\text{L}}' = 0, \\[1ex]
\begin{bmatrix}
q_{\text{R}} \\[1ex]
q_{\text{R}}'
\end{bmatrix} = 
\begin{bmatrix}
\mspc\frac{5}{2} &  -\frac{3}{2} \mspc \\[\matspace]
\mspc\frac{5}{3} & -\frac{5}{3} \mspc
\end{bmatrix}
\begin{bmatrix}
\bar{q} \\
q_{\text{L}}
\end{bmatrix},
\end{array}
\qquad
\begin{array}{l}
\text{if }\left(\xi_{\text{I}}^{*} = {+1}\right)\text{ and } \left(\sigma q_{\text{R}}' \leq 0\right), \quad q_{\text{R}}' = 0, \\[1ex]
\begin{bmatrix}
q_{\text{L}} \\[1ex]
q_{\text{L}}'
\end{bmatrix} = 
\begin{bmatrix}
 \mspc 5 &  -4 \\[\matspace]
-10 & \mspc 10   
\end{bmatrix}
\begin{bmatrix}
\bar{q} \\
q_{\text{R}}
\end{bmatrix}
\end{array}
\end{eqnarray}
Following these final modifications to the edge-values and edge-slopes, the resulting slope-limited PQM interpolant is guaranteed to enforce exact monotonicity.

\section*{References}

\bibliographystyle{elsarticle-harv}
\bibliography{references}

\end{document}